\documentclass[11pt,a4paper]{article}
\usepackage{amsmath,amssymb,bbold,epsfig}
\usepackage{amsfonts}
\usepackage{slashed}
\usepackage{dsfont}
\usepackage{cite}
\usepackage{graphicx,color}
\usepackage{bbm}
\usepackage{bm}
\usepackage[normalem]{ulem}
\usepackage[usenames,dvipsnames]{xcolor}
\definecolor{lightgray}{gray}{0.90}
\usepackage[colorlinks=true,linkcolor=blue,citecolor=green,urlcolor=cyan,
bookmarks=true,bookmarksopen=true,pdfpagemode=None,pdfstartview=FitH]{hyperref}
\newcommand{\appsection}{\addtocounter{section}{1}\setcounter{equation}{0}
                         \renewcommand{\thesection}{\Alph{section}}
}
\renewcommand{\theequation}{\arabic{equation}}
\newcommand{\be}{\begin{equation}}
\newcommand{\ee}{\end{equation}}
\newcommand{\bea}{\begin{eqnarray}}
\newcommand{\eea}{\end{eqnarray}}

\begin{document}

\title{\vspace{-2cm}
\vglue 1.3cm
\Large \bf
Coherent scattering and macroscopic coherence: Implications for 
neutrino, dark matter and axion detection}
\author{
{Evgeny~Akhmedov\!\footnote{Also at the National Research Centre
Kurchatov Institute, Moscow, Russia}~\thanks{Email: \tt
akhmedov@mpi-hd.mpg.de}\,,
\;Giorgio~Arcadi\thanks{Email: \tt arcadi@mpi-hd.mpg.de}\,, 
\;Manfred~Lindner\thanks{Email: \tt lindner@mpi-hd.mpg.de}\; and 
Stefan~Vogl\thanks{E-mail: \tt stefan.vogl@mpi-hd.mpg.de}
\vspace*{3.5mm}
} \\
{\normalsize\em
Max-Planck-Institut f\"ur Kernphysik, Saupfercheckweg 1,}\\
{\normalsize\em
69117 Heidelberg, Germany
\vspace*{0.15cm}}
}
\date{}

\maketitle 
\thispagestyle{empty} 
\begin{abstract} 
We study the question of whether coherent neutrino scattering can 
occur on macroscopic scales, leading to a significant increase of the 
detection cross section. We concentrate on radiative neutrino 
scattering on atomic electrons (or on free electrons in a conductor). 
Such processes can be coherent provided that the net electron recoil 
momentum, i.e.\ the momentum transfer from the neutrino minus the 
momentum of the emitted photon, is sufficiently small. The radiative 
processes is an attractive possibility as the energy of the emitted 
photons can be as large as the momentum transfer to the electron 
system and therefore the problem of detecting extremely low energy 
recoils can be avoided.  The requirement of macroscopic coherence 
severely constrains the phase space available for the scattered 
particle and the emitted photon. We show that in the case of the 
scattering mediated by the usual weak neutral current and charged 
current interactions this leads to a strong suppression of the 
elementary cross sections and therefore the requirement of 
macroscopic coherence results in a reduction rather than an increase 
of the total detection cross section.  However, for the $\nu e$ 
scattering mediated by neutrino magnetic or electric dipole moments 
coherence effects can actually increase the detection rates.  Effects 
of macroscopic coherence can also allow detection of neutrinos in 100 
eV -- a few keV energy range, which is currently not accessible to 
the experiment. A similar coherent enhancement mechanism can work for  
relativistic particles in the dark sector, but not for the 
conventionally considered non-relativistic dark matter.

\end{abstract} 
\vspace{0.2cm} 
\vspace{0.3cm}

\newpage

{\color{black}
\tableofcontents}

\section{\label{sec:intro}Introduction}
Recently, the COHERENT collaboration has reported the first observation of 
coherent elastic neutrino--nucleus scattering \cite{COH1,COH2}, a process 
predicted over forty years ago \cite{Freed1,Freed2}.  
This observation completed the standard-model picture of neutrino interactions 
with nucleons and nuclei and opened up a new window to probe physics beyond the 
standard model and nuclear structure; it also has important implications 
for astrophysics. 
Very recently, the CONUS collaboration has reported the first experimental 
indication of coherent elastic neutrino-nucleus scattering with reactor 
antineutrinos \cite{CONUS1}.
Coherence of the process implies that the total cross section is proportional 
to the squared number of the target particles rather than to their number; 
as a result, for the first time it became possible to observe neutrinos 
with a hand-held detector rather than with ton- or kiloton-scale ones  
-- a spectacular achievement indeed.  One then naturally wonders if it is 
possible to achieve coherence of neutrino detection on scales that are 
larger than the nuclear scale, such as atomic or even macroscopic scales, 
leading to a further significant increase of the detection cross sections. 
This would also be of great interest for detecting Dark Matter (DM) particles 
which are currently being actively looked for. 

Coherent neutrino scattering on atoms \cite{nuAtom1,nuAtom2,nuAtom3} 
has a two-fold advantage. First, the scattering would occur not just on 
nucleons inside the nucleus but also on atomic electrons, and 
the increased number of scatterers would mean additional enhancement of the 
detection cross section. Second, within the standard model, $\nu_e e$ 
scattering proceeds through both charged-current (CC) and neutral-current (NC) 
weak interactions, whereas the $\nu_{\mu,\tau}e$ scattering is mediated only 
by neutral currents. Therefore, coherent neutrino--atom scattering would be 
sensitive to neutrino flavour and thus could potentially be used for studying 
neutrino oscillations. This is in contrast with the already observed coherent 
elastic neutrino--nucleus scattering proceeding only through neutral-current 
interactions which are flavour blind. 

The problem with coherent neutrino-atom scattering 
is that the atomic recoil energies would be very small and extremely difficult 
to measure. Indeed, coherence requires the momentum transfer to the scatterer 
$|\vec{q}|$ to be smaller than or at most of the order of the inverse radius 
of the scatterer.  
It is only under this condition that it will be impossible to find out on 
which constituent of the target particle has the neutrino scattered, and the 
neutrino waves scattered from the different constituents will be in phase with 
each other, which are the necessary conditions for coherent scattering. For 
neutrino--atom scattering, this would imply  
\be
|\vec{q}|\lesssim (\mbox{a few}~a_B)^{-1}\sim 1~{\rm keV}\,,
\label{eq:cond1}
\ee
where $a_B\simeq 0.53{\rm \AA}$ is the Bohr radius. For an atom with 
the atomic number $A\sim 100$ the recoil energy would then be  
\be
E_{rec}\simeq \frac{\vec{q}\,^2}{2m_A}\sim 10^{-5}~{\rm eV}\,, 
\label{eq:recoil1}
\ee
about eight orders of magnitude below the currently achieved sensitivity. 
Measuring such small recoil energies 
presents a 
formidable experimental challenge and, if possible at all, would 
probably require 
new technologies.  

\subsection{\label{sec:macrcoh}Macroscopic coherence?}
How about scattering with coherence on macroscopic scales? Clearly,  
this would require measuring even much smaller recoil energies and so 
does not look practical. It is interesting, however, to inquire what could be 
the increase of the detection cross sections if such measurements were 
possible, leaving for the moment the detection problem aside. For an estimate, 
we will be assuming coherence on the target length scale of $\sim 1$ cm and the 
target mass $m_t\sim 1\,g$. 
The total cross section of the elementary neutrino elastic 
scattering process (i.e.\ of the scattering on a single target particle) 
with non-relativistic 
target particle recoil is  
$\sigma_0\simeq (G_F^2/\pi)\omega^2$, where $G_F$ is the Fermi constant and 
$\omega$ is the energy of the incident neutrino. 
To achieve macroscopic coherence, we need momentum transfers satisfying 
$|\vec{q}|\le q_0\sim 
(1~{\rm cm})^{-1}$.%
\footnote{Note that by $q_0$ we mean the maximum allowed value of $|\vec{q}|$ 
and not the time component of the 4-vector $q$.}
However, the energies of neutrinos we normally deal with 
are many orders of magnitude larger than this value, 
and so are the typical momentum transfers. From the kinematics of elastic 
scattering it follows that $\vec{q}\,^2\simeq 2\omega^2(1-\cos\theta)$, where 
$\theta$ is the neutrino scattering angle; therefore, to achieve 
macroscopic coherence  
one has to restrict neutrino scattering 
to nearly forward directions:   
\be
1-\cos\theta\le \frac{ \vec{q}_0\,^2}{2\omega^2}\ll 1\,.
\label{eq:forw1}
\ee
This means a severe restriction of the phase space accessible to 
the final-state neutrino, which, in turn, leads to a strong suppression 
of the corresponding elementary cross section: 
\be
\sigma_0\simeq \frac{G_F^2}{\pi}\omega^2~~\longrightarrow~~
\frac{G_F^2}{2\pi}q_0^2\,. 
\label{eq:crsec2}
\ee
However, in order  
to find the cross section per one target particle one has to multiply 
the elementary cross section 
(\ref{eq:crsec2}) by the number of particles that contribute coherently to 
the scattering process, i.e.\ by the number of particles in the coherent 
volume $L_0^3\sim 1/q_0^3$. 
As a result, the cross section per target particle will be proportional to 
$1/q_0\propto N^{1/3}$, where $N$ is the number of scatterers in the target 
and we have assumed that the coherent volume is comparable with the total 
volume of the target. 
Thus, by going to smaller $q_0$ one could 
increase the detection cross section.
\footnote{Obviously, one cannot go to the limit $q_0\to 0$, as the above 
estimates are only valid for coherence volumes $1/q_0^3$ not exceeding the 
total volume of the detector.} 
The total cross section obtained by summing over all the scatterers in the 
target will then scale as $N^{4/3}$, i.e.\ the cross section 
increase due to the coherence effects is $\sim N^{1/3}$. While this is 
much smaller than an extra factor of $\sim N$ one could naively expect, it 
still would mean a very strong enhancement of the detection cross section.  

The problem is, of course, that the recoil energies are too small to be 
detected. For $q_0\sim (1~{\rm cm})^{-1}\simeq 2\times 10^{-5}$ eV and the 
total mass of particles in the coherent volume $m_t\sim 1\,g$, one finds  
$E_{rec}\sim \vec{q_0}^2/2m_t\sim 10^{-43}~{\rm eV}$, the quantity which 
is not going to be ever measured. To give just one reason for that, in order 
to measure recoil energy of this magnitude, one needs an energy resolution of 
at least the same order of magnitude, $\delta E\sim E_{rec}$. By time-energy 
uncertainty relation, the duration of the measurement process $\delta t$ 
should then exceed $\sim 10^{27} s$, which is about 10 orders of magnitude 
larger than the age of the Universe. 
To summarize, macroscopic coherence holds and the cross section becomes very 
large only for neutrino scattering in a very narrow forward cone, which 
corresponds to unmeasurably small recoil energies of the target particles.   

As is seen from the above discussion, 
one reason why it is difficult to 
achieve macroscopic coherence in neutrino scattering processes is that one 
usually measures the recoil energy of the target particles, which for small 
recoils is suppressed compared to the recoil momentum by a very small factor 
$\sim q_0/2m_t$. The same applies, of course, to experiments on direct DM 
detection.%
\footnote{
Macroscopic coherence is, however, readily achieved in experiments on light, 
$X$-ray or neutron scattering from macroscopic targets because what is 
detected are the scattered particles and not the recoil of the target.} 

Is it possible to overcome this obstacle by somehow making use of the recoil 
momentum rather than the recoil energy? In that case no extra suppression 
factor $q_0/2m_t$ would be there. One such possibility was suggested in the 
1980s by Joseph Weber \cite{weber1,weber2,weber3}.

\subsection{\label{sec:weber}Weber's approach and structure factors}
Weber suggested to 
detect neutrinos through their coherent scattering on crystals in torsion 
balance experiments. This approach combines two interesting ideas.  First, 
as the force coincides with momentum transfer per unit time, the force 
neutrinos impinge on a crystal is directly related to the momentum transfer 
to the target rather than to the recoil energy. As discussed above, this is 
a very desirable feature. Second, when the expected recoil energy of the 
individual atoms $E_{rec}=\vec{q}\,^2/2m_A$ 
is below the Debye temperature of the crystal $T_D$, 
the recoil momentum is with high probability given to the crystal as a whole 
rather than to the individual atoms, 
similarly to what happens in the M\"{o}ssbauer effect. Indeed, the recoil-free 
fraction is approximately given by \cite{Moessb} 
\be
f\simeq \exp\left\{-\frac{E_{rec}}{T_D}\left(\frac{3}{2}+\frac{\pi^2 T^2}{T_D^2}
\right)\right\}
\label{eq:rec-free}
\ee 
where $T$ is the crystal temperature.%
\footnote{
This formula is valid for harmonic crystals in the limit 
$T\ll T_D$. For a general $T$, the second term in the round brackets in the 
exponent should be replaced by $6(T/T_D)^2\int_0^{T_D/T} dx x/(e^x-1)$. 
}
For $E_{rec}\ll T_D$ the quantity $f\simeq 1$, i.e.\ the momentum is 
transferred to the crystal as a whole with probability close to 1. 
For typical M\"{o}ssbauer crystals $T_D\sim 10$ keV, and the 
condition $E_{rec}\ll T_D$ is easily satisfied even for neutrinos in 
the 10 MeV energy range. Weber asserted that, since in this case it 
is impossible to find out on exactly which atom the neutrino had 
scattered, the contributions of different scatterers should add up 
coherently, leading to macroscopic coherence and a very strong 
enhancement of the detection cross section.

He developed a theoretical approach to describe neutrino coherent 
scattering on crystals and obtained encouraging results. He then performed 
experiments with solar neutrinos, reactor antineutrinos and a radioactive 
neutrino source and in all three cases reported positive results, in reasonable 
agreement with his theoretical expectations. 

These results were met with scepticism, and were strongly criticized by a 
number of authors. It was pointed out that the same ideas 
applied to the $X$-ray \cite{Xray} and neutron \cite{neutrons} scattering on 
crystals would lead to unrealistically large cross sections in direct 
contradiction with experiment. In Refs.~\cite{casella1,butler,aharonov,
smith1,lipkin,trammell} the theoretical approach 
of~\cite{weber1,weber2,weber3} was criticized. It was concluded that the 
effect had been overestimated by about 24 orders of magnitude. Finally, 
subsequent torsion balance experiments on neutrino-crystal scattering with 
sensitivities much higher than the sensitivity of the Weber's device have 
reported null result \cite{null1,null2}. 
   
So, what went wrong with Weber's ideas? The absence of recoil of the 
individual atoms, which was the main ingredient of his approach, is 
necessary for macroscopic coherence, but is not sufficient. 
It is also necessary that the neutrino waves scattered from different 
centers be in phase with each other. 
The amplitudes of particle scattering on a group of scattering centers rather 
than on a single center should contain the relevant structure factors, which 
describe the relative phases of the amplitudes corresponding to different 
scatterers. For elastic neutrino scattering 
the structure factor is given by 
\be
F(\vec{k}-\vec{k}')
=\sum_{i=1}^N e^{i(\vec{k}-\vec{k}')\vec{r}_i}\,, 
\label{eq:F1}
\ee
where $\vec{k}$ and $\vec{k}'$ are the momenta of the incident and scattered 
neutrinos, 
$\vec{r}_i$ is the coordinate of the $i$th 
scatterer and $N$ is the total number of scatterers in the target.  
Introducing the number density of scatterers $\rho(\vec{r})=\sum_i
\delta^3(\vec{r}-\vec{r}_i)$, one can rewrite the structure factor 
(\ref{eq:F1}) 
in the familiar form-factor form
\be
F(\vec{q})=\int d^3 r e^{i\vec{q}\vec{r}}\rho(\vec{r})\,,  
\label{eq:F2}
\ee 
where $\vec{q}=\vec{k}-\vec{k}'$ is the momentum transfer to the target. 

The structure factors 
are crucial to the issue of coherence of the scattering process, i.e.\ to 
the question of whether the amplitudes of neutrino scattering on different 
target particles should be added coherently. 
While the exact form of these factors depend 
on the specific target utilized in the 
experiment, the fully coherent and completely incoherent regimes can be studied 
in a rather general way. 
Indeed, 
the squared modulus of the transition amplitude contains the factor 
\be
|F(\vec{q})|^2
=\sum_{i,j=1}^N e^{i\vec{q}(\vec{r}_i-\vec{r}_j)}\,.  
\label{eq:F3}
\ee
If the momentum transfer 
$\vec{q}$ satisfies the condition 
\be
\max\limits_{i,j}\{|\vec{q}(\vec{r}_i-\vec{r}_j)|\}\simeq 
|\vec{q}|L \ll 1\,, 
\label{eq:cond2}
\ee
(where $L$ is a linear size of the target), one can replace all the phase 
factors under the sum in eq.~(\ref{eq:F3}) by unity, which gives 
$|F(\vec{q})|^2=N^2$.%
\footnote{As discussed above, in the cross section this dependence reduces to 
$\sim N^{4/3}$ if neutrino scattering has to be restricted to nearly forward 
directions in order to achieve sufficiently small momentum transfers.}
In this case neutrinos scattered from different 
constituents of the target are in phase with each other.  
In the opposite limit $|\vec{q}|L \gg 1$ only the 
diagonal ($i=j$) terms in the sum survive, and one finds  
$|F(\vec{q})|^2\simeq N$, i.e.\ we obtain the usual dependence of the total 
cross section on the number of the target particles. This corresponds to 
incoherent neutrino scattering.

For scattering on crystals, yet another possibility of having macroscopically 
coherent effects exists, namely, when the phase differences 
$\vec{q}(\vec{r}_i-\vec{r}_j)$ in eq.~(\ref{eq:F3}) are integer multiples of 
$2\pi$. This leads to the well known Bragg condition for diffraction on 
crystals, 
\be
2d\sin\vartheta = n\lambda\,,
\label{eq:Bragg}
\ee
where $d$ is the interplanar distance in the crystal, $\vartheta$ is the angle 
between the neutrino momentum and the atomic plane (the scattering angle being 
$\theta=2\vartheta$), $\lambda=2\pi/|\vec{k}|$ and $n$ is an integer. 
Just like for $X$-ray diffraction on crystals, the 
intensity of the scattered neutrino wave 
in the directions of the Bragg maxima is $\propto N^2$. It is noticeably 
different from zero in narrow cones around the Bragg directions, 
with the corresponding solid angles $\Delta \Omega\propto N^{-2/3}$, and is 
practically zero outside these cones.  Thus, the intensity of the scattered 
neutrino wave around each Bragg maximum is $\propto N^{4/3}$ \cite{LL8}. 
Since the scattered neutrinos are not detected, the quantity that is in 
principle measurable is the crystal recoil momentum, or the force impinged 
on the crystal. For a given direction of the momenta of the incident neutrinos 
with respect to the crystal atomic planes and $n\ne 0$, 
eq.~(\ref{eq:Bragg}) selects the neutrino energy that satisfies 
the Bragg condition.%
\footnote{For $n=0$ the Bragg condition is satisfied for all neutrino energies. 
However, it corresponds to forward scattering in which there is no momentum 
transfer from neutrinos to the crystal.}
As the Bragg maxima have finite widths, neutrinos in finite energy intervals 
$\Delta \omega$ will actually experience Bragg diffraction; 
these intervals are, however, very small and scale as  
$1/L\propto N^{-1/3}$. As a result,
the overall momentum transfer to the crystal scales as 
$N^{4/3}\times N^{-1/3}=N$, just like for the scattering on amorphous bodies 
\cite{LL8}.

Thus, scattering on crystals unfortunately does not give any advantage for 
neutrino detection, and one is back to consider the condition in 
eq.~(\ref{eq:cond2}). Since it was {\sl not} satisfied in Weber's experiments, 
macroscopic coherence could not be achieved.%
\footnote{
Note that Weber actually did consider the structure factors, but 
evaluated them incorrectly \cite{butler}.
}
\footnote{
When the preliminary results of this work were presented at CERN 
neutrino platform week in January 2018, we were informed by P. Huber that 
some of the considerations presented in Sections~\ref{sec:macrcoh} 
and \ref{sec:weber} had appeared earlier in the unpublished 
(but not classified) Jason Report by Callan, Dashen and Treiman 
\cite{Report}. We thank Patrick Huber for this 
comment and for sending us a scanned copy of \cite{Report}.
}
%

\subsection
{\label{sec:radnusc}Our approach: Radiative neutrino scattering on electrons}
In the present paper we consider a different realization of the idea of 
employing the momentum transfer to the target rather than the recoil energy 
of the target particle -- radiative neutrino scattering on atomic electrons or 
on free electrons in a conductor:
\be
\nu+e\to \nu+e+\gamma\,.
\label{eq:proc1}
\ee
In this case the emitted photon rather than the recoil electron is detected, 
and the photon energy $\omega_\gamma$ can be as large as the neutrino 
momentum transfer 
$|\vec{k}-\vec{k'}|$.
Most importantly, the momentum transfer itself (and so also $\omega_\gamma$) 
need not be small in order to ensure macroscopic coherence of the process. 
What has to be small 
($\lesssim L_0^{-1}$ where $L_0$ is the macroscopic 
length scale of the coherent volume) is the net recoil momentum of the target 
particle, which is the difference between the momentum transfer from the 
neutrinos $\vec{k}-\vec{k'}$ and the momentum $\vec{k}_\gamma$ carried away by 
the photon. This can happen even when 
$|\vec{k}-\vec{k'}|$ 
and $|\vec{k}_\gamma|=\omega_\gamma$ are both large compared to 
$L_0^{-1}$. 
The above points directly follow from the expression for the structure factor 
in the case of the process (\ref{eq:proc1})
(cf. eq.~(\ref{eq:F1})),  
\be
F(\vec{k}-\vec{k}'-\vec{k}_\gamma) 
=\sum_{i=1}^N e^{i(\vec{k}-\vec{k}'-\vec{k}_\gamma)\vec{r}_i}\,.
\label{eq:F4}
\ee
The condition for the scatterers within a volume $\sim L_0^3$\, to contribute 
coherently is $|\vec{k}-\vec{k}'-\vec{k}_\gamma|L_0\lesssim 1$.

Note that in both radiative and 
elastic scattering cases, momentum 
conservation implies that the argument of the structure factor coincides 
with the momentum $\vec{p}\,'$ of the recoil electron. 
This has a simple physical interpretation. As was discussed above, coherence 
requires $|\vec{p}\,'|L\ll 1$. Since the uncertainty of the magnitude of 
momentum 
cannot be much larger than the momentum itself, we also have in this case 
$\delta|\vec{p}\,'|L\ll 1$. The Heisenberg uncertainty relation then means 
that the coordinate uncertainty of the recoiling electron $\delta x$ 
exceeds the size of the target, that is, one cannot identify which electron 
the neutrino was scattered off. 
The requirement $|\vec{p}\,'|L\ll 1$ also ensures that the neutrino 
waves scattered from all the electrons within the volume $L^3$ are in phase 
with each other.  
These are precisely the conditions of coherence of 
the contributions of different individual electrons to the amplitude of the 
process. 

\subsubsection
{\label{sec:previous}Previous studies}

The radiative neutrino scattering on electrons (\ref{eq:proc1}) was first 
considered by Lee and Sirlin back in 1964~\cite{leesirlin} and since then has 
been studied by many authors (see, e.g., \cite{zhizhin1,mourao,
loebstark,bahcgould,
websehg,buccella,
bernab}). To the best of our knowledge, 
only two studies \cite{loebstark,bahcgould} concern the issue of 
macroscopic coherence of the process. 
In \cite{loebstark} it was suggested to use radiative neutrino scattering 
on free electrons in a conductor in order to detect cosmic background 
neutrinos. It was argued that macroscopic coherence of the process can be 
achieved, leading to measurable photon production cross sections. 
These results have been criticized in \cite{bahcgould}, where a crucial 
flaw of \cite{loebstark} was pointed out. It was demonstrated that, as 
neutrino impact pushes the conduction electrons deeper inside the target, 
the excess positive ion charge on its surface creates a restoring force 
which pulls the electrons back. As a result, the cross section of coherent  
radiative neutrino scattering gets suppressed by a factor $(\omega/
\omega_{p})^4$, where $\omega_p=(n_e e^2/m_e)^{1/2}\sim 10$ eV is the 
plasma frequency. For cosmic background neutrinos $(\omega/\omega_{p})^4\sim 
10^{-20}$, which makes the process completely unobservable. 

It is actually not difficult to understand the reason for this drastic 
suppression of the photon production cross section. 
Photon radiation in process (\ref{eq:proc1}) is due to the time dependent 
dipole (and in general higher multipole) moments induced by the 
neutrino scattering on the electrons of the target. In the very long 
wavelength limit, when the energy transfer to the system (and so also the 
frequency of the induced radiation) is small compared to the characteristic 
frequencies of the system, the induced moments are small and the photon 
radiation is strongly suppressed.
This situation is very similar to the one encountered when comparing the cross 
section of the Rayleigh scattering (photon scattering on bound electrons in 
atoms) 
to that of the Thomson scattering (scattering of photons on free electrons). 
In the classical limit the two 
cross sections are related by \cite{jackson} 
\footnote{The accurate quantum mechanical formula is more complicated and 
depends sensitively on the atomic structure, see Section~\ref{sec:atomic} 
below. However,  the limits of 
large and small $\omega$ are reproduced by the classical formula 
(\ref{eq:RayThom}) correctly.}
\be
\sigma_R\simeq\frac{\omega^4}{(\omega_{at}^2-\omega^2)^2}\sigma_T\,,
\label{eq:RayThom}
\ee   
where $\omega_{at}$ is a characteristic atomic frequency. In the limit 
$\omega\gg \omega_{at}$ the two cross sections coincide, i.e.\ the photon 
scattering on atomic electrons proceeds as if the electrons were free. 
In the opposite limit $\omega\ll \omega_{at}$ one finds $\sigma_R
\simeq(\omega/\omega_{at})^4\sigma_T\ll \sigma_T$. This is the famous 
$\omega^4$ law which is responsible for the blue color of the sky. The 
$\sim\omega^4$ suppression of the cross section of radiative scattering of 
cosmic background neutrinos on conduction electrons found in 
\cite{bahcgould} is of exactly the same nature.%
\footnote{In Ref.~\cite{bahcgould} 
this suppression was incorrectly interpreted as being due to the electric 
neutrality of the target. 

For neutrino scattering on a charged conductor the restoring force on 
the electrons accelerated by the neutrino impact would still be 
there, and would be due to both the pull from the positive ions and 
push from the excess electrons. As a result, the $\omega^4$ 
suppression would still be present. This is quite analogous to the 
situation with photon scattering on atomic systems, where the 
scattering on charged ions exhibits at $\omega\ll \omega_{at}$ the 
same $\omega^4$ suppression as the scattering on neutral atoms 
\cite{carney,fisak}.
}

\subsubsection{\label{sec:above}Radiative scattering with 
$\omega\gtrsim \omega_{\rm char}$  and phase space constraints}

In the present paper we shall consider neutrino detection through coherent 
radiative neutrino scattering on atomic electrons or on free electrons in 
a conductor. We will be assuming the energies of the incident neutrinos 
to be higher than the corresponding 
characteristic atomic frequencies $\omega_{at}$ or plasma frequencies 
$\omega_p$. This will allow the momenta of the emitted photons to 
exceed $\omega_{at}$ and $\omega_p$, thus avoiding the 
$\omega^4$ suppression of the cross sections discussed above. We will 
concentrate on the situations when the momentum carried away by the emitted 
photon nearly compensates the momentum transfer to electrons from neutrinos, 
leading to very small net recoil momenta of the target 
electrons. As discussed at the beginning of 
Section~\ref{sec:radnusc}, this will result in macroscopic coherence 
of the detection process, while completely avoiding the problem of 
measuring extremely small recoil energies of the target.

There is a price to pay, however. The requirement 
$\vec{k}-\vec{k'}\simeq \vec{k}_\gamma$ puts 
a stringent constraint on the phase space volume accessible to the 
final-state particles, and in general also on the amplitude of the 
process. This should lead to a suppression of the cross section of 
the individual process, just like in the case of elastic neutrino 
scattering discussed in Section~\ref{sec:macrcoh} (see 
eq.~(\ref{eq:crsec2})). It has to be seen if the enhancement of the 
cross section due to macroscopic coherence can overcome this 
suppression, as it is the case for the
elastic neutrino scattering (which, 
however, is unobservable because of the vanishingly small recoil energies). 
In the present paper we study this issue in detail. 

We find that for radiative neutrino scattering mediated by the 
standard NC and CC weak interactions macroscopic coherence can occur, 
but only at the expense of severe restriction of the kinematics of 
the process, resulting in the net suppression rather than enhancement 
of the total cross section. In contrast to this, for radiative 
neutrino scattering mediated by neutrino magnetic or electric dipole 
moments the net effect is an enhancement of the cross section per 
target electron compared to that for the elastic scattering, though 
only for the kinetic energies of electron recoil in the elastic process 
exceeding $\sim 100$ keV. In addition, coherent radiative $\nu e$ 
scattering due to neutrino magnetic or electric dipole moments could 
potentially allow detection of neutrinos of very low energies, which are 
currently not accessible to the experiment. The mechanism we consider here 
is, unfortunately, not operative for conventional (non-relativistic) DM 
particle candidates; however, it could work for relativistic particles that 
may exist in the dark sector. 

\subsection{\label{sec:structure}The structure of this paper}

The paper is organized as follows. In Section~\ref{sec:neutrino} we consider 
the radiative neutrino scattering on free non-relativistic electrons, both 
without any additional kinematic constraints and assuming that the electron 
recoil momentum is limited from above by a small value $p_0$, allowing for 
macroscopic coherence of the process. In Section~\ref{sec:neutrinoWeak} we 
discuss the radiative 
neutrino scattering on free electrons mediated by the usual NC and CC weak 
interactions, whereas in Section~\ref{sec:numagn} we study the case when the 
$\nu e$ scattering is mediated by the neutrino magnetic or electric dipole 
moments. In Section~\ref{sec:DM} 
we briefly discuss the question of whether macroscopic coherence could be 
realized and the same enhancement mechanism could work for direct DM 
detection and conclude that for conventional DM this is not possible 
(mainly for kinematic reasons). In Section~\ref{sec:axions} we consider
coherent radiative axion-photon conversion due to scattering of relativistic
axions on electrons (radiative inverse Primakoff effect). 
In Section~\ref{sec:atomic} we discuss atomic binding effects in the case 
when radiative scattering takes place on electrons in an atom rather than 
on free electrons. We demonstrate that these effects can be neglected 
in the cases of interest to us. In Section~\ref{sec:coheff} we use the cross 
sections obtained in Sections~\ref{sec:neutrino} and \ref{sec:axions} 
to consider the effects of macroscopic coherence on radiative neutrino 
scattering and axion-photon conversion processes 
and the question of whether it can increase the detection cross sections. 
We summarize and discuss our results in Section~\ref{sec:disc}. 
Some technical details of our calculations are given in 
the Appendices. The kinematics of radiative $2\to 3$ scattering 
is considered in Appendix A, whereas Appendix B describes calculations 
of the integrals over the 3-body phase space. The expressions for the squared 
matrix elements for the processes studied in the paper are collected in 
Appendix C.

\section{\label{sec:neutrino} Radiative neutrino scattering on electrons}

We shall consider the process 
\be
\nu(k)+e(p)\to \nu(k')+e(p')+\gamma(k_\gamma)
\label{eq:proc1a}
\ee 
in the rest frame of the initial electron. 
Here
\be
k=(\omega, \vec{k})\,,\quad
p=(m, \vec{0})\,,\quad
k'=(\omega', \vec{k}')\,,\quad
p'=(E_{p'}, \vec{p}\,')\,,\quad
k_\gamma=(\omega_\gamma, \vec{k}_\gamma)\,
\label{eq:4mom}
\ee
are the 4-momenta of the incident neutrino, initial-state electron, scattered 
neutrino, final-state electron and emitted photon, respectively. In this 
section we consider radiative neutrino scattering 
on free non-relativistic electrons; possible effects 
of atomic binding will be discussed in Section~\ref{sec:atomic}. Eventually, 
we will be interested in coherent radiative neutrino scattering on a 
macroscopic lump of electrons, which we will assume to be unpolarised, i.e.\ 
to have zero total spin. This allows us to simplify the problem by neglecting 
the electron spin, i.e.\ to consider 
neutrino scattering on a ``spinless electron'' -- a particle with the 
electron's charge and mass but zero spin. Neutrinos are assumed to be 
ultra-relativistic, so that the neutrino mass can be neglected both in the 
kinematics of the process and in calculating transition matrix elements.

We shall consider neutrino-electron scattering mediated either by the usual 
NC and CC weak interactions or by neutrino magnetic (or electric) dipole 
moments. In each case we calculate the cross section first 
allowing all the final-state momenta to span the full ranges allowed by  
4-momentum conservation,%
\footnote{
Except that for $\omega_\gamma$-integrated cross sections an infrared 
cutoff $\omega_0$ will be introduced for the photon energies, see below.}
and then restricting the net recoil momentum of the electron to satisfy 
$|\vec{p}\,'|\le p_0$, where $p_0$ is small compared to the maximum value 
of $|\vec{p}\,'|$ allowed by the kinematics of process. 
We will need the cross sections with such a kinematic restriction when 
considering macroscopic coherence effects in Section~\ref{sec:coheff}.

\subsection{\label{sec:neutrinoWeak} Weak interactions induced radiative 
process}

In calculating the cross section of 
radiative neutrino scattering (\ref{eq:proc1a}) on ``spinless electron'' 
we take into account only the vector current part of the weak NC and CC 
interactions of electrons since the axial-vector current does not 
contribute to neutrino scattering on zero-spin targets. 
\begin{figure}[ht]
\begin{tabular}{ccc}
\includegraphics[width=4.8cm,height=3.8cm]{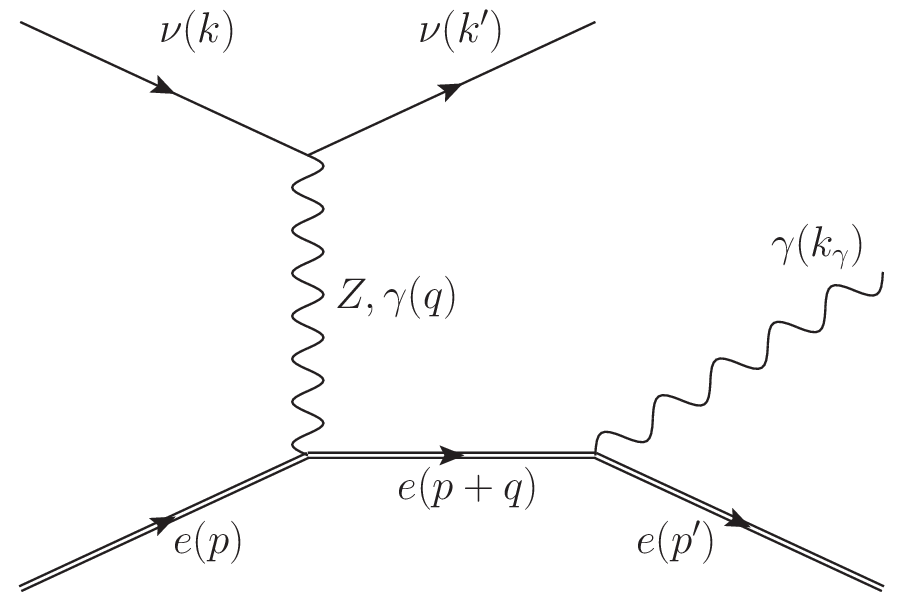}&
\includegraphics[width=4.8cm,height=3.8cm]{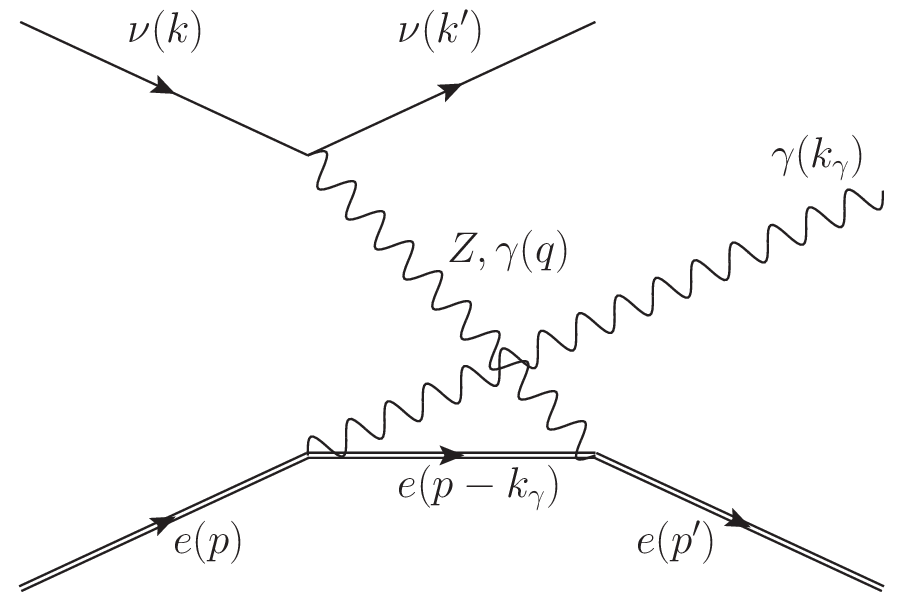}&
\includegraphics[width=4.8cm,height=3.8cm]{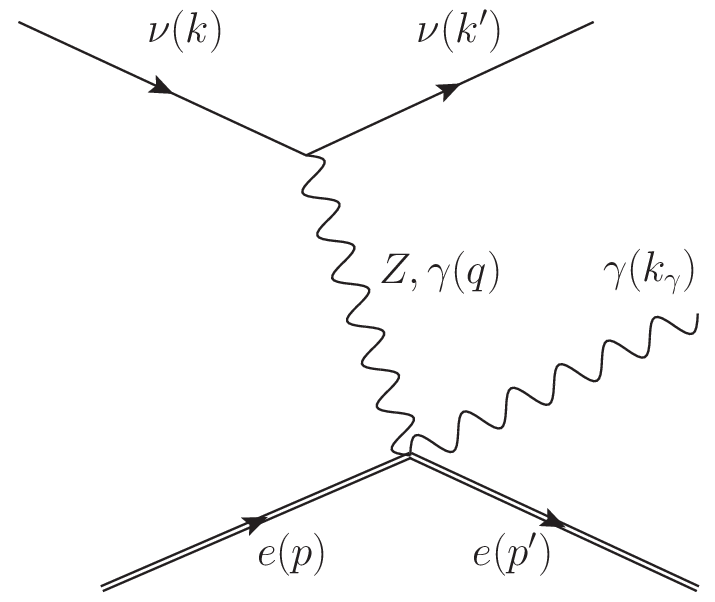}
\end{tabular}
\caption{\label{fig:FD1}
\small 
Leading order Feynman diagrams for radiative neutrino scattering 
(\ref{eq:proc1}) on a ``spinless electron''. Shown are the diagrams for 
$\nu e$ scattering mediated by $Z^0$ exchange 
(NC weak interaction) or photon exchange (for scattering due to the neutrino 
magnetic or electric dipole moments). The latter case will be considered 
in Section~\ref{sec:numagn}. For the CC weak interaction contributions, see 
the text.  
}
\end{figure}
The amplitude of the weak interaction induced radiative neutrino 
scattering on a ``spinless electron'' can be written as 
\be
{\cal M}_w=-i\frac{G_F}{\sqrt{2}}g_V e \epsilon_\mu^*(k_\gamma)
Q^{\mu\alpha}
j_\alpha\,.
\label{eq:M1}
\ee
Here $g_V$ is the vector weak coupling constant, 
$e$ is the electron charge, $\epsilon^\mu(k_\gamma)$ is the polarization 
vector of the produced photon and 
\be
j_\alpha=\bar{u}(k')\gamma_\alpha(1-\gamma_5)u(k)\,
\label{eq:j}
\ee
is the matrix element of the neutrino current. To leading order in 
electroweak interaction the tensor $Q^{\mu\alpha}$ is given by 
\be
Q^{\mu\alpha}=Q^{\mu\alpha}(p,k;\,p',k',k_\gamma)\equiv 
\left\{
\frac{(2p'+k_\gamma)^\mu(2p+k-k')^\alpha}{2p'\cdot k_\gamma}
-\frac{(2p-k_\gamma)^\mu[2p'-(k-k')]^\alpha}{2p\cdot k_\gamma}
-2g^{\mu\alpha}\right\}.
\vspace*{2.5mm}
\label{eq:Q1}
\ee
The subscript $w$ at ${\cal M}$ stands for neutrino-electron scattering  
due to the weak interactions; for $\nu_\mu e$ and $\nu_\tau e$ scattering 
the interaction is mediated by the weak neutral current, whereas for the 
$\nu_e e$ scattering both neutral and charged currents contribute. 
For NC induced radiative $\nu e$ scattering, the leading order amplitude 
is described by the diagrams 
of Fig.~\ref{fig:FD1}. The three terms in $Q^{\mu\alpha}$ correspond to the 
three diagrams shown there. For CC induced radiative scattering of $\nu_e$  
on a ``spinless electron'', one cannot directly draw diagrams similar to 
those in Fig.~\ref{fig:FD1}, as the vertex $W e\nu$ connecting spin 1, 0 
and 1/2 fields does not exist. Instead, one should consider the scattering 
on the ``standard'' spin 1/2 electron described by the left and middle 
diagrams of Fig.~\ref{fig:FD1} with the $Z^0$ boson line replaced by the 
$W^\pm$ boson one and the electron and neutrino lines in the final state 
interchanged. For unpolarised target electrons, 
in the limit of non-relativistic electron recoils the electron spin becomes 
relatively unimportant, and the corresponding CC amplitude   
again has the form (\ref{eq:M1}) with $Q^{\mu\alpha}$ given by 
eq.~(\ref{eq:Q1}).%
\footnote{To arrive at this result one has to make use of the Fierz 
transformation and consider unpolarised 
electrons in the limit when their recoil energy is non-relativistic in the 
rest frame of the initial-state electrons. Note that, as we are interested  
in coherent effects, the summation over the electron spin states should be 
done at the amplitude level.}
Note that $Q^{\mu\alpha}$ satisfies the gauge invariance conditions 
\be
k_{\gamma\mu}Q^{\mu\alpha}=Q^{\mu\alpha}(k-k')_\alpha=0\,.
\label{eq:gauge1}
\ee

Thus, expressions (\ref{eq:M1}) and (\ref{eq:Q1}) adequately describe 
the amplitude of process (\ref{eq:proc1}) for non-relativistic electrons, 
with both NC and CC contributions properly taken into account.   	
The coupling constant $g_V$ is given by 
\be
g_V=\left\{\begin{array}{ll}2\sin^2\theta_W+\frac{1}{2}\,,& \nu=\nu_e
\vspace*{2mm}\\
2\sin^2\theta_W-\frac{1}{2}\,,& \nu=\nu_\mu,\, \nu_\tau\end{array}
\right.\!. 
\label{eq:WE7}
\ee

We now proceed to calculate the cross sections, first without  
constraining $|\vec{p}\,'|$. 
For the double and single differential cross sections 
one finds  
\be
\frac{d^2\sigma_w}{d\omega_\gamma d\cos\theta_\gamma}=
\frac{G_F^2 g_V^2 e^2}{(2\pi)^3}\frac{1}{m_e^2}\cdot\frac{1}{3}
\frac{(\omega-\omega_\gamma)^2}{\omega_\gamma}\big\{
(\omega^2-4\omega\omega_\gamma)(1-\cos^2\theta_\gamma)+2(\omega^2
+2\omega_\gamma^2)
\big\}
\,,
\label{eq:WE1}
\vspace*{2.0mm}
\ee
\be
\frac{d\sigma_w}{d\omega_\gamma}=\frac{G_F^2 g_V^2 e^2}{(2\pi)^3}\frac{1}{m_e^2}
\cdot\frac{8}{9}
\frac{(\omega-\omega_\gamma)^2}{\omega_\gamma}\big\{
(\omega-\omega_\gamma)^2+\omega^2+2\omega_\gamma^2
\big\}
\,.
\label{eq:WE2}
\vspace*{2mm}
\ee
Here $m_e$ is the electron mass and $\theta_\gamma$ is the angle between the 
momentum of the emitted photon and that of the incident neutrino. 
Because of the usual infrared divergence, in order to calculate the 
$\omega_\gamma$-integrated cross section one has to introduce a lower cutoff 
for the energy of the emitted photon $\omega_{\gamma\rm min}\equiv 
\omega_0$. In our case a natural choice 
of $\omega_0$ follows from the requirement that the photon energy exceed 
the characteristic frequency of the target system, $\omega_{at}$ for scattering 
on atomic electrons or $\omega_p\sim 10$ eV for scattering on free electrons in 
a conductor. As discussed in Section~\ref{sec:radnusc}, this will allow one to 
avoid the $\sim\omega^4$ suppression of the cross section. 

Integration of (\ref{eq:WE2}) over $\omega_\gamma$ yields 
\be
\sigma_w(\omega_\gamma>\omega_0)=\frac{G_F^2 g_V^2 e^2}{(2\pi)^3}\frac{1}{m_e^2}
\cdot\frac{16}{9}\omega^4 \big\{\ln(1/x)-\frac{41}{24}+3x-\frac{9}{4}x^2 
+\frac{4}{3}x^3-\frac{3}{8}x^4
\big\}
\,,
\label{eq:WE3}
\ee
where
\be
x\equiv
\omega_0/\omega\,.
\label{eq:x}
\ee
For $\omega_0\ll \omega$ eq.~(\ref{eq:WE3}) gives 
\be
\sigma_w(\omega_\gamma>\omega_0)\simeq\frac{G_F^2 g_V^2 e^2}{(2\pi)^3}
\frac{1}{m_e^2}\cdot\frac{16}{9}\omega^4 \Big[\ln(\omega/\omega_0)-
\frac{41}{24}\Big]
\,.
\label{eq:WE3a}
\ee

Next, we constrain the momentum of the final-state electron by requiring 
$|\vec{p}\,'|\le p_0$, where $p_0$ is small compared to $|\vec{p}\,'|_{\max}$ 
allowed by 4-momentum conservation. The kinematics of the process in this case 
is considered in Appendix B. As shown there, for a given $\omega_\gamma$ the 
photon emission angle $\theta_\gamma$ is now constrained by 
\be
0\le 1-\cos\theta_\gamma\le \frac{p_0^2+2p_0(\omega-\omega_\gamma)}
{2\omega\omega_\gamma}\,.
\label{eq:allowed1}
\ee
The smallness of $p_0$ implies that the photons are emitted in nearly 
forward direction. From 
the kinematics of the process it follows that the same is true for the 
scattered neutrino. 
In the leading order in $p_0$ we obtain 
\be
\frac{d\sigma_w}{d\omega_\gamma}=\frac{G_F^2 g_V^2 e^2}{(2\pi)^3}\frac{p_0^4}
{4m_e^2}
\frac{\omega^2+(\omega-\omega_\gamma)^2}{\omega^2\omega_\gamma}
\,.
\vspace*{3mm}
\label{eq:WE4}
\ee
Here the integration over $\cos\theta_\gamma$ was performed in its allowed 
range given in eq.~(\ref{eq:allowed1}). 

The cross section for the emission of photons with energies 
$\omega_\gamma\ge \omega_{\gamma\rm min}\equiv \omega_0$ reads 
\be
\sigma_w(\omega_\gamma>\omega_0)=\frac{G_F^2 g_V^2 e^2}{(2\pi)^3}\frac{p_0^4}
{2m_e^2}
\Big\{\ln(1/x)-\frac{3}{4}+x-\frac{1}{4}x^2 \Big\}
\,.
\label{eq:WE5}
\ee
For $\omega_0\ll \omega$ one can retain only the first two terms in the 
curly brackets.  

The cross sections in eqs.~(\ref{eq:WE4}) and (\ref{eq:WE5}) 
scale as the fourth power of $p_0$; a factor $p_0^3$ is expected from 
the phase space volume of the process with 
the electron recoil momentum 
constrained by $|\vec{p}\,'|\le p_0$ 
(see Appendix B), and one more power of $p_0$ comes from the squared modulus 
of the transition matrix element of the process.

\subsection{\label{sec:numagn} 
Radiative scattering and the neutrino magnetic dipole moment}

Let us now consider the radiative neutrino scattering process 
(\ref{eq:proc1}) in the case when the neutrino-electron scattering is 
mediated by the photon exchange due to neutrino magnetic or electric dipole 
moments. In what follows we will for definiteness discuss  the case of 
neutrino magnetic dipole moment $\mu_\nu$. We will comment on the general 
case at the end of this subsection. 

The amplitude of the process corresponds to the diagrams in Fig.~1 in which 
the intermediate vector boson connecting the neutrino and electron lines is 
the photon.%
\footnote{We ignore the possibility that the 
final-state photon is emitted 
from the neutrino 
line, as this would be a process of higher order in the very small neutrino 
magnetic moment $\mu_\nu$.}
In this case one can expect some kinematic enhancement compared to the usual 
weak NC and CC induced processes considered in the previous subsection. 
Indeed, we are interested in the kinematic region in which the momentum 
$\vec{k}_\gamma$ carried away by the photon nearly coincides with the 
momentum transfer from the neutrino, $\vec{q}=\vec{k}-\vec{k}'$. In the 
regime of small net recoil momentum of the electron, the same is true for 
the corresponding energies: $\omega_\gamma\simeq \omega-\omega'$. This 
means that the 4-momentum of the virtual photon $q=k-k'$ nearly coincides 
with that of the final-state photon, $k_\gamma$. As the produced photon is 
on the mass shell, 
the virtual photon is nearly on the mass shell, and its propagator 
should lead to an enhancement of the amplitude of the process. 

The transition matrix element of the neutrino magnetic moment induced radiative 
scattering process on a ``spinless electron'' is 
\be
{\cal M}_m=-ie^2\frac{\mu_\nu}
{q^2}
\epsilon_{\mu}^*(k_\gamma)Q^{\mu\alpha}\tilde{j}_\alpha \,,
\quad\qquad q\equiv k-k'\,.
\label{eq:M2}
\ee
Here 
\be
\tilde{j}_\alpha=\bar{u}(k')\sigma_{\alpha\beta}q^\beta u(k)=
(k+k')_\alpha \bar{u}(k')u(k)\,,
\label{eq:Lalpha}
\vspace*{2mm}
\ee
where we have used the Gordon identity and took into account that neutrinos 
are treated as massless particles. As before, the kinematic regime of 
non-relativistic electron recoil is considered. Without constraining 
$|\vec{p}\,'|$, for the double and single differential cross sections we 
find 
\be
\frac{d^2\sigma_m}{d\omega_\gamma d\cos\theta_\gamma}=
\frac{\mu_\nu^2 e^4}{(2\pi)^3}\frac{1}{4m_e^2}\cdot
\frac{(\omega-\omega_\gamma)^2}{\omega_\gamma}\big(3-\cos^2\theta_\gamma
\big)
\,,
\label{eq:MM1}
\ee
\be
\frac{d\sigma_m}{d\omega_\gamma}=
\frac{\mu_\nu^2 e^4}{(2\pi)^3}\frac{1}{m_e^2}\cdot\frac{4}{3}
\frac{(\omega-\omega_\gamma)^2}{\omega_\gamma}
\,.
\vspace*{3mm}
\label{eq:MM2}
\ee
The cross section for the emission of photons with energies 
$\omega_\gamma\ge \omega_{\gamma\rm min}\equiv \omega_0$ is   
\be
\sigma_m(\omega_\gamma>\omega_0)=
\frac{\mu_\nu^2 e^4}{(2\pi)^3}\frac{1}{m_e^2}\cdot\frac{4}{3}\omega^2 
\big\{\ln(1/x)-\frac{3}{2}+2x-\frac{1}{2}x^2 \big\}\,.
\label{eq:MM3}
\ee
For $\omega_0\ll \omega$ this equation gives 
\be
\sigma_m(\omega_\gamma>\omega_0)=
\frac{\mu_\nu^2 e^4}{(2\pi)^3}\frac{1}{m_e^2}\cdot\frac{4}{3}\omega^2 
\Big\{\ln(\omega/\omega_0)-\frac{3}{2}\Big\}
\,.
\label{eq:MM4}
\ee

Next, we again constrain the momentum of the final-state electron by 
requiring $|\vec{p}\,'|\le p_0$. Integrating over the $\theta_\gamma$ in 
the allowed range given in eq.~(\ref{eq:allowed1}), we find, to leading 
order in \vspace*{0.5mm} $p_0$,
\be
\frac{d\sigma_m}{d\omega_\gamma}=
\frac{\mu_\nu^2 e^4}{(2\pi)^3}\frac{1}{m_e^2}\cdot\frac{1}{6}
\frac{(\omega-\omega_\gamma)p_0^3}{\omega\omega_\gamma^2}
\,,
\label{eq:MM5}
\vspace*{1mm}
\ee
\be
\sigma_m(\omega_\gamma>\omega_0)=
\frac{\mu_\nu^2 e^4}{(2\pi)^3}\frac{1}{m_e^2}\cdot\frac{1}{6}\frac{p_0^3}
{\omega_0}\,
\Big\{1-x-x\ln(1/x)\Big\}
\,.
\label{eq:MM6}
\ee
For $\omega_0\ll \omega$ this gives 
\be
\sigma_m(\omega_\gamma>\omega_0)\simeq
\frac{\mu_\nu^2 e^4}{(2\pi)^3}\frac{1}{m_e^2}\cdot\frac{1}{6}\frac{p_0^3}
{\omega_0}=
\frac{\mu_\nu^2 \alpha^2}{\pi}\frac{1}{m_e^2}\cdot\frac{1}{3}\frac{p_0^3}
{\omega_0}
\,.
\vspace*{1mm}
\label{eq:MM7}
\ee
Interestingly, in this approximation the cross section is essentially 
independent of the incident neutrino energy $\omega$, except that 
$\omega_0$ should satisfy $\omega_0<\omega$.  

The cross sections (\ref{eq:MM5})-(\ref{eq:MM7}) 
increase with decreasing minimum photon energy $\omega_0$.  
Recall, however, that the photon energy 
cannot be too small: It should exceed the characteristic frequency 
($\omega_{at}$ for neutrino scattering and $\omega_p$ for scattering 
on free electrons in a conductor) in order to avoid the $\omega^4$ 
suppression. 

As discussed at the beginning of this subsection, the cross section of 
the neutrino magnetic moment induced process exhibits for small 
$p_0$ a kinematic enhancement due to the propagator of the 
virtual photon being close to its pole. The enhancement, 
however, turns out to be relatively mild: the cross sections 
(\ref{eq:MM5})-(\ref{eq:MM7}) scale as $p_0^3$, 
which is to be compared with the $p_0^4$ dependence 
found in Section~\ref{sec:neutrinoWeak}.   

We have considered here radiative neutrino scattering process 
(\ref{eq:proc1}) induced by the neutrino magnetic dipole moment $\mu_\nu$. 
In general, neutrinos may have both the magnetic 
and electric dipole moments, which, in addition, are matrices in flavour 
space. One can take this into account by replacing in the expression for 
the transition amplitude the quantity 
$\mu_\nu$ by $\tilde{\mu}_{\alpha\beta}\equiv 
(\mu_\nu+i\epsilon_\nu)_{\alpha\beta}$, where  
$(\epsilon_\nu)_{\alpha\beta}$ is the matrix of neutrino electric dipole 
moments. Such an amplitude will then describe the transition of 
$\nu_\alpha$ to a neutrino $\nu_\beta$ which may be 
of the same or different flavour. 
As the final-state neutrino is not detected, in calculating the cross 
section of the process one has to sum over $\beta$. For the 
ultra-relativistic neutrinos we confine ourselves to, this amounts to 
replacing in the expressions for the cross sections   
$\mu_\nu^2\to \sum_\beta|\tilde{\mu}_{\alpha\beta}|^2$.

\section{\label{sec:DM} DM detection through radiative coherent scattering?}

It would be interesting to extend the above considerations to detection of 
other particles, such as DM. Unfortunately, the mechanism of enhancement of 
the detection cross section through macroscopic coherence considered here for 
neutrinos would not work for non-relativistic projectiles. 
The reason is actually 
mostly kinematic. Macroscopic coherence requires tiny 
net recoil momenta $\vec{p}\,'$ of the target electrons. It is easy to see 
that for non-relativistic projectiles vanishing $\vec{p}\,'$ is excluded 
by energy-momentum conservation (see Appendix A). Small non-zero values of 
$\vec{p}\,'$ are allowed, but only for extremely soft emitted photons, 
$\omega_\gamma \ll |\vec{p}\,'|$ (or $\omega_\gamma \lesssim |\vec{p}\,'|$ 
in the case of moderately relativistic projectiles). 
As discussed above, the cross sections of radiative scattering on electrons 
get a very strong $\sim\omega_\gamma^4$ suppression in this case.   

As the conventionally discussed DM particles are supposed to be 
non-relativistic, the detection enhancement mechanism 
considered here will not be operative for them.%
\footnote{In Ref.~\cite{kouvaris} it was suggested to use the radiative 
coherent scattering on nuclei to detect DM particles, but the issue of 
macroscopic coherence has not been addressed there.}  
It may, however, work for detection of relativistic particles that may 
exist in the dark sector. 
 
\section{\label{sec:axions} Coherent detection of relativistic axions}

The enhancement mechanism considered in the present paper could also
work for detection of relativistic axions (such as e.g.\ axions from the sun). 
The detection process is 
\be
a(k)+e(p)\to e(p')+\gamma(k_1)+\gamma(k_2)\,.
\label{eq:proc1b}
\ee 
In this case there are two photons in the final state, with 4-momenta 
$k_1$ and $k_2$. 
\begin{figure}[t]
\begin{tabular}{ccc}
\includegraphics[width=4.8cm,height=3.8cm]{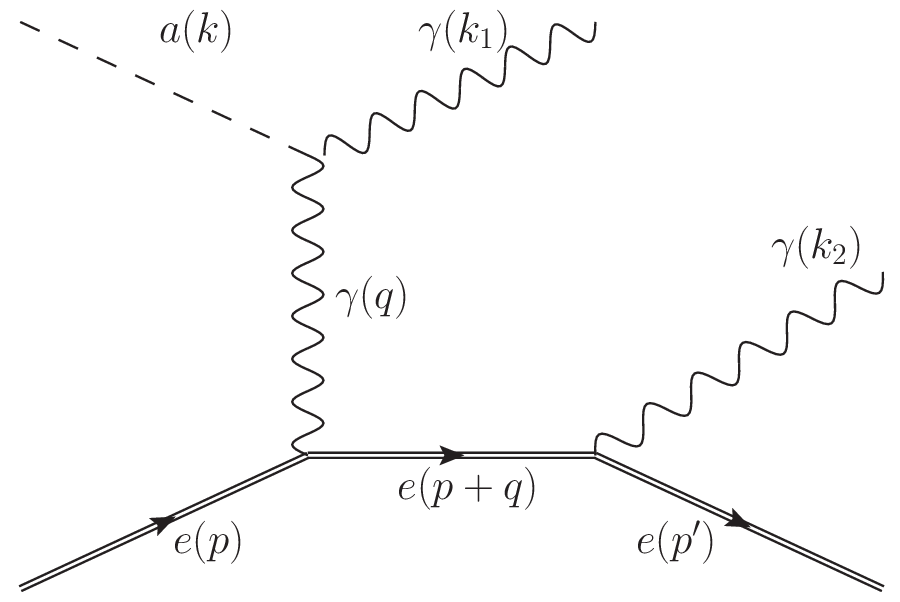}&
\includegraphics[width=4.8cm,height=3.8cm]{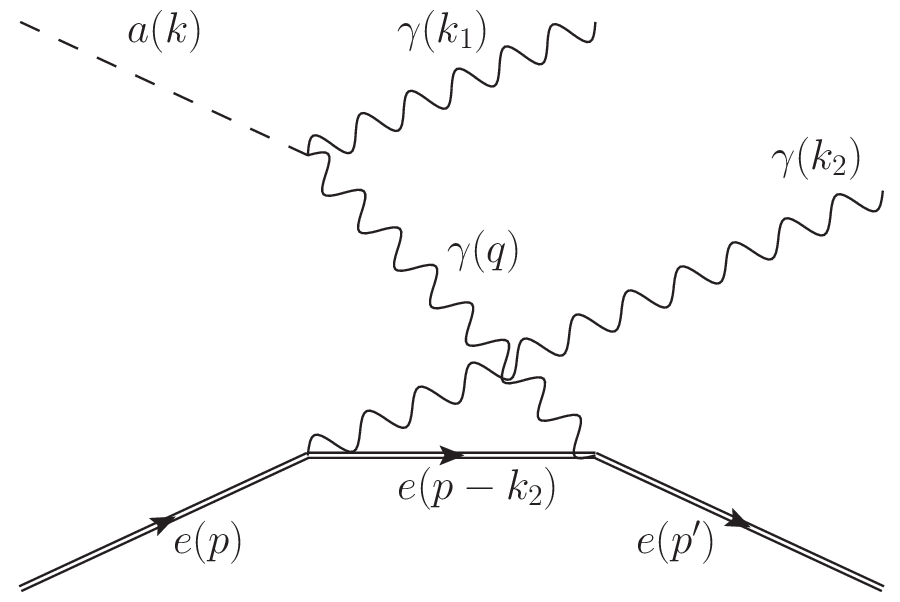}&
\includegraphics[width=4.8cm,height=3.8cm]{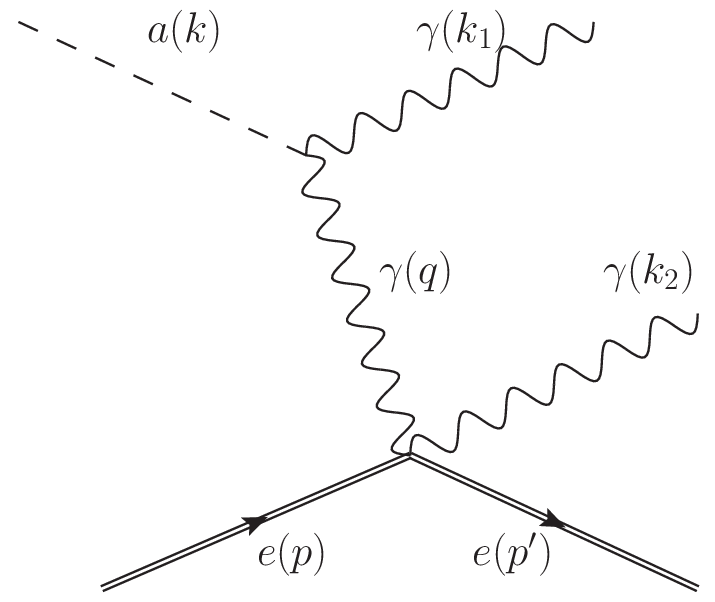}
\end{tabular}
\caption{\label{fig:FD2}
\small 
{ 
Leading order Feynman diagrams contributing to radiative axion-photon 
conversion on electrons (\ref{eq:proc1b}) through the inverse Primakoff 
mechanism. 
The diagrams with interchanged photon 4-momenta $k_1$ and $k_2$ 
should be added.  
}}
\end{figure}  

The mechanism of the process (\ref{eq:proc1b}) that we consider is the 
radiative inverse Primakoff effect. As before, we will be interested in the 
coherent interaction of the projectile particle with a group of electrons of 
zero total spin, which allows us to adopt the approximation of a ``spinless 
electron''. The leading order diagrams contributing to the process are shown 
in Fig.~\ref{fig:FD2}. Note that the non-radiative inverse Primakoff effect on 
charged particles is incoherent, as the 
contributions of scattering on electrons and positively charged nuclei 
cancel each other in the case of electrically neutral targets for low 
momentum transfers \cite{Raffelt1}. Coherent axion-photon conversion is, 
however, possible in external magnetic fields \cite{Sikivie:1983ip}. We do not 
consider Compton-like photon production (or double photon production) 
because the direct axion-electron coupling is spin dependent and so it 
cannot be coherently enhanced in the case of unpolarized targets.
 
We consider the $a\gamma\gamma$ interaction Lagrangian 
\be
{\cal L}=\frac{1}{4}g_{a\gamma\gamma}  a F_{\mu\nu}\tilde{F}^{\mu\nu}.
\label{eq:Lagr}
\ee
The amplitude of the process corresponding to the diagrams of Fig.~2 can be 
written in the form similar to (\ref{eq:M2}):
\be
{\cal M}_a=-ie^2 g_{a\gamma\gamma}
\left\{\frac{1}{q_1^2}  \epsilon_{\mu}^*(k_1)
Q^{\mu\alpha}(p,k;\,p',k_1,k_2) 
\hat{j}_{2,\alpha} +\frac{1}{q_2^2} \epsilon_{\mu}^*(k_2)
Q^{\mu\alpha}(p,k;\,p',k_2,k_1) 
\hat{j}_{1,\alpha}  \right\}.
\label{eq:M3}
\ee
Here 
\be
\hat{j}_{i,\alpha}=\varepsilon_{\rho\beta\sigma\alpha}{k}_i^{\rho}
\epsilon^{*\beta}(k_i)q^\sigma\,,\qquad\quad q_i\equiv k-k_i, 
\label{eq:current}
\vspace*{2mm}
\ee
and the tensor $Q^{\mu\alpha}(p,k;\,p',k_1,k_2)$ was defined 
in eq.~(\ref{eq:Q1}). We will neglect the axion 
mass $m_a$ in all the calculations, except that we will keep the 
$g_{a\gamma\gamma}$ coupling constant (which is usually assumed to be 
proportional to $m_a$) finite. 
As before, we calculate all the cross sections in the regime of 
non-relativistic electrons, first without any additional constraints on 
$|\vec{p}\,'|$. For detection of one of the two emitted photons the double 
and single differential cross section are 
\[
\frac{d^2\sigma_a}{d\omega_1 d\cos\theta_1}=\frac{g_{a\gamma \gamma}^2 
e^4}{(2 \pi)^3}\frac{1}
{96 m_e^2\, \omega\omega_1 (\omega -\omega_1 )} 
\left\{4\omega^4-16 \omega^3 \omega_1 +25 \omega^2 \omega_1^2-17 \omega 
\omega_1^3+8\omega_1 ^4\right.
\vspace*{-0.8mm}
\]
\be
\qquad\qquad\qquad\qquad
\left.+\omega \omega_1\left[(3 \omega ^2-9 \omega \omega_1 +10 \omega_1^2)
\cos\theta_1
+3(\omega -\omega_1)^2\cos^2\theta_1\right]\right\}, 
\label{eq:ax0}
\vspace*{2.5mm}
\ee
\be
\frac{d\sigma_a}{d\omega_1}=\frac{g_{a\gamma\gamma}^2 e^4}{(2\pi)^3}\
\frac{1}{ 
12m_e^2\, \omega \omega_1 (\omega-\omega_1)}\left(\omega^4-\frac{15}{4}
\omega^3 \omega_1+\frac{23}{4} 
\omega^2 \omega_1^2-4 \omega \omega_1^3+2 \omega_1^4\right).
\label{eq:ax1}
\vspace*{2mm}
\ee
To find the total cross section of the process we have to integrate 
(\ref{eq:ax1}) over $\omega_1$ in the interval $[\omega_0,\, \omega-\omega_0]$. 
The upper integration limit is $\omega-\omega_0$ rather than $\omega$ because 
we now have two photons in the final state, whose energies are related by the 
energy conservation condition $\omega_1+\omega_2=\omega$ and must both be 
above the infrared cutoff $\omega_0$. For the total cross section we then find 
\be 
\sigma_a= \frac{e^4 g_{a\gamma 
\gamma}^2}{(2\pi)^3}\frac{\omega^2}{6 m_e^2}\left\{ 
\mbox{ln}\left(\frac{1}{x}-1\right) -\frac{41}{24} + \frac{15}{4} x - 
x^2+ \frac{2}{3}x^3 \right\} 
\ee 
with $x=\omega_0/ \omega$. Note that $x$ must satisfy $x<1/2$ 
as otherwise energy conservation would force one of the photon energies to 
be below $\omega_0$ in violation of our assumption. 

Next, we consider the case in which $|\vec{p}\,'|$ is constrained from above 
by a small value $p_0$. For the differential and total cross sections 
we find 
\be
\frac{d\sigma_a}{d\omega_1}=\frac{g_{a\gamma\gamma}^2 e^4}{(2\pi)^3}
\frac{p_0^3}{48 m_e^2 \omega^2}
\left(\frac{\omega_1}{\omega-\omega_1}+\frac{\omega-\omega_1}{\omega_1}\right)
\label{eq:ax2}
\ee
and 
\be
\sigma_a
=\frac{g_{a\gamma\gamma}^2 e^4}{(2\pi)^3}\,
\frac{p_0^3}{24 m_e^2 \omega}\,
\left[\ln\left(\frac{1}{x}-1\right)-1+2 x\right]. 
\label{eq:ax3}
\ee

\section{\label{sec:atomic}Effects of atomic binding}
In 
Sections~\ref{sec:neutrino} and~\ref{sec:axions} we considered radiative
neutrino scattering and radiative axion-photon conversion on free
electrons. This is suitable for conduction electrons in metals; 
however, for scattering on atomic electrons in dielectrics the effects of 
atomic binding should in general be taken into account. We shall 
show now that for the 
kinematic regime of interest to us, when the net recoil momentum of the 
electron is small and at same time the neutrino or axion 
energy satisfies $\omega\gg \omega_{at}$, the atomic effects can be neglected 
and the results found in Section~\ref{sec:neutrino} apply. 

To demonstrate this, let us first note that for radiative scattering on free 
electrons, in the regime of small $|\vec{p}\,'|$ the contribution of the 
first two terms in the expression for $Q^{\mu\alpha}$ (\ref{eq:Q1})
is small, and the main contribution comes from 
the third term, corresponding to the right 
diagram of Fig.~1.%
\footnote{Indeed, for small electron recoil momenta 
the terms in $Q^{\mu\alpha}$ 
proportional to ${p'}^\mu p^\alpha$ and  to $p^\mu {p'}^\alpha$ nearly 
cancel each other, whereas the terms $\propto (k-k')^\alpha$ are subleading 
in the case of non-relativistic electrons and weak NC and CC mediated 
neutrino-electron scattering 
(their contributions vanish exactly for neutrino magnetic moment mediated 
$\nu e$ scattering as well as for axion-photon conversion). 
The terms $\propto k_\gamma^\mu$ do not contribute by gauge invariance. Note 
that for calculations in the Coulomb gauge in the rest frame of the 
initial-state electron the whole second term in $Q^{\mu\nu}$ does not 
contribute to the amplitude, and the contribution of the first term is small 
because of $p'\approx p$.    
}
The same holds true when atomic effects are taken into account:
in the kinematic region of interest to us the analogues of the first two 
terms in $Q^{\mu\alpha}$ are small, and the main contribution comes form 
the analogue of the third term, which is largely insensitive to the effects 
of atomic structure (see below).
This is fully analogous to what happens for elastic scattering of photons 
on atoms when the photon energy is much higher than the characteristic 
atomic frequencies $\omega_{at}$.  
As discussed in Section~\ref{sec:radnusc}, in this limit the cross section 
essentially coincides with that of photon
scattering on free electrons. This can be readily seen from the 
expression for the amplitude of elastic photon-atom scattering. For 
non-relativistic electrons, the leading order 
amplitude in the Coulomb gauge is proportional to \cite{atampl,kane}
\be
-\frac{1}{m}\sum_n\! \left\{ \frac{\langle i|e^{-i\vec{k}_f\vec{r}} 
\vec{\rm p}
\vec{\epsilon}_{f}^{\hspace*{0.6mm}*} 
|n\rangle \langle n| e^{i\vec{k}_i\vec{r}} 
\vec{\rm{p}}\vec{\epsilon}_i 
|i\rangle}{E_n-E_i-\omega_i-i\varepsilon}+
\frac{\langle i|e^{i\vec{k}_i\vec{r}}
\vec{\rm{p}}\vec{\epsilon}_i 
|n\rangle \langle n| e^{-i\vec{k}_f\vec{r}} 
\vec{\rm{p}}
\vec{\epsilon}_{f}^{\hspace*{0.6mm}*}
|i\rangle}{E_n-E_i+\omega_i-i\varepsilon}
\right\}\!+
(\vec{\epsilon}_{f}^{\hspace*{0.6mm}*}\!\cdot\vec{\epsilon}_i)
\langle i| e^{i(\vec{k}_i-\vec{k}_f)\vec{r}}|i\rangle.
\label{eq:atom1}
\ee
Here 
$\vec{\rm{p}}=-i\vec{\nabla}$ is the 3-momentum operator, $\vec{k}_i$ and 
$\vec{k}_f$ are the momenta of the incident and scattered photons, 
$\vec{\epsilon}_i$ and $\vec{\epsilon}_f$ are their polarizations vectors, 
and the sum is over the intermediate atomic states. 
In eq.~(\ref{eq:atom1}) we have taken 
into account that for elastic scattering on a heavy system 
$\omega_f=|\vec{k}_f|$ coincides with $\omega_i=|\vec{k}_i|$. 
For $\omega_i\ll \omega_{at}$ all three terms in (\ref{eq:atom1}) are of the 
same order of magnitude and nearly cancel each other, leading to the 
$\sim \omega_i^4$ suppression mentioned in Section~\ref{sec:radnusc}; 
however, in the regime $\omega_i\gg \omega_{at}$ that is of interest to us, 
the first two terms in (\ref{eq:atom1}) are small compared to the third term 
and to a good accuracy can be neglected. Moreover, for spherically symmetric 
atomic states $|i\rangle$ they tend to cancel each other.%
\footnote{Indeed, using the closure property of the atomic states and 
commuting the factors $e^{i\vec{k}_i\vec{r}}$, $e^{-i\vec{k}_f\vec{r}}$ with 
the momentum operator, for the sum of the first two terms in (\ref{eq:atom1}) 
one finds in this limit \,$-(1/m\omega_i)
\epsilon_f^{*l}\epsilon_i^{s}e^{i(\vec{k}_i-\vec{k}_f)\vec{r}}
\langle i|k_i^l {\rm p}^s+k_f^s {\rm p}^l|i\rangle$, which vanishes for 
spherically symmetric states $|i\rangle$. 
As we are interested in coherent scattering on a group of atoms,  
by $|i\rangle$ one should actually understand the ground state of such a 
system. The cancellation then happens also in the case when this state is 
spherically symmetric (i.e.\ has zero total angular momentum), even if the 
ground states of the individual atoms are not.} 
The remaining term,  
$(\vec{\epsilon}_{f}^{\hspace*{0.6mm}*}\!\cdot\vec{\epsilon}_i)
\langle i| e^{i(\vec{k}_i-\vec{k}_f)\vec{r}}|i\rangle$, in general depends on 
the electron charge distribution in the state $|i\rangle$. 
For $|\vec{k}_i-\vec{k}_f|\ll R_{at}^{-1}$ it is actually independent of the 
atomic structure, and for photon scattering on a single atom reduces to 
$Z(\vec{\epsilon}_{f}^{\hspace*{0.6mm}*}\!\cdot\vec{\epsilon}_i)$, where 
$Z$ is the total number of the atomic electrons. This corresponds to coherent 
elastic photon-atom scattering. If the more stringent condition 
$|\vec{k}_i-\vec{k}_f|\ll L^{-1}$ is satisfied where $L$ is the linear size 
of the target, the scattering 
on all electrons in the target is coherent. Otherwise, one would need to 
take into account structure factors describing electron distribution in the 
target, as discussed in Section~\ref{sec:weber}.

Similar arguments apply to radiative neutrino 
scattering on atoms. Note that in this case one has to replace in 
eq.~(\ref{eq:atom1}) 
\be
\vec{k}_i\to \vec{k}-\vec{k}'\,, \qquad \omega_i\to\omega-
\omega'\,,\qquad \vec{k}_f\to \vec{k}_\gamma\,,\qquad 
\omega_f\to\omega_\gamma\,, 
\qquad\vec{\epsilon}_i \to \vec{j}   
\quad {\rm and}\quad{\vec{\rm p}}\vec{\epsilon}_i \to \rm p\cdot \!j\,,
\label{eq:replace}
\ee
where ${\rm p}^\mu=i\partial^\mu$ is the 4-momentum operator and 
$j^\mu=(j^0,\,\vec{j}\,)$ is the relevant matrix element of the neutrino 
current. The condition $|\vec{k}_i-\vec{k}_f|\ll R_{at}^{-1}$ is then replaced 
by $|\vec{k}-\vec{k}'-\vec{k}_\gamma|=|\vec{p}\,'|\ll R_{at}^{-1}$, which 
we always assume to be satisfied with a large margin when discussing 
macroscopically coherent effects. With minor modifications
related to the presence of two photons in the final state, the same argument
applies also to radiative axion-photon conversion on atoms. 

It should be noted that for $\omega\gg \omega_{at}$ inelastic scattering 
with ionization or excitation of atoms typically dominates, while the 
processes in which the atom remains in its initial state are only important 
for nearly forward scattering. This is, however, exactly the 
case we are interested in. The fact that the probability of the radiative 
scattering without excitation or ionization of the target atoms is small is 
already taken into account by 
the suppression of the individual cross sections which we found upon 
constraining the electron recoil momentum by 
$|\vec{p}\,'|\lesssim p_0\sim 10^{-5}$ eV.

\section{\label{sec:coheff}Coherent effects and the cross sections}
Let us now assess the effects of macroscopic coherence on the cross sections
of neutrino and axion detection processes (\ref{eq:proc1a}) and
(\ref{eq:proc1b}). 

As discussed in Sections~\ref{sec:weber} and \ref{sec:radnusc}, in order to 
take possible macroscopic coherence effects into account one has to 
multiply the elementary amplitude of the process 
by the relevant structure factor (such as (\ref{eq:F1}) or (\ref{eq:F4})). 
The structure factor depends on the target used in the experiment, 
and the calculated cross section will therefore also be   
target-dependent. However, 
simple estimates of the effects of macroscopic coherence 
can be obtained in a rather general way as follows. 

Assume that all the scatterers contained in some volume of a linear size 
$L_0$ within the target contribute to the cross section coherently; for 
this to occur, the net recoil momentum of the scatterer $\vec{p}\,'$ must 
satisfy $|\vec{p}\,'|\lesssim p_0\sim 2\pi L_0^{-1}$. 
The coherent volume $L_0^3$ can in principle range from just the volume 
per one scatterer (no coherence) to the total volume of the target $L^3$ 
(complete coherence). To assess the coherence effects one can first 
calculate the elementary cross section of the process with the constraint 
$|\vec{p}\,'|\le p_0$ imposed. In calculating such constrained elementary 
cross sections the corresponding structure factors 
can be replaced by unity. To find the cross section 
per target particle with coherence effects taken into account one 
would then have to multiply the constrained elementary cross section by the 
number of scatterers in the coherent volume $L_0^3\simeq (2\pi/p_0)^3$.%
\footnote{
Let the number of scatterers within one coherent volume be $N_0$, and the 
number of coherent volumes in the target be $k$. The total number of 
scatterers in the target is $N=kN_0$. 
If $\sigma_0$ is 
the elementary cross section of the process, the cross section corresponding 
to scattering on all the target particles contained within one coherent 
volume is $\sigma_0 N_0^2$. The total cross section is 
$\sigma_0 N_0^2\times k 
=\sigma_0 N_0 N$.  
The cross section per one target particle is then $\sigma_0 N_0$, as stated. 
In the fully coherent case ($N_0=N$) and completely incoherent case 
($N_0=1$) the total cross sections are 
$\sigma_0 N^2$ and $\sigma_0 N$, respectively, and 
the corresponding cross sections per target particle are $\sigma_0 N$ and 
$\sigma_0$.}
The choice of the recoil momentum cutoff $p_0$ (i.e.\ of the linear size of 
$L_0$ of the coherent volume) would then have to be optimized, within the 
range allowed by the kinematics of the process and the geometry of the 
experiment, by maximizing the resulting cross section. 

In doing this, one should not forget the issue of observability of the 
process, which may be fully coherent but 
completely unobservable. For example, as discussed in 
Section~\ref{sec:macrcoh}, for 
elastic neutrino scattering the optimization requires to choose for the 
maximum recoil momentum (denoted $q_0$ there) the 
smallest possible value $q_0\sim L^{-1}$, but the scattering will then 
be unobservable due to the vanishingly small recoil energy of the target 
particles (see the discussion around eqs.~(\ref{eq:forw1}) and 
(\ref{eq:crsec2})). No such problems arise for radiative processes 
discussed in the present paper.

We shall now estimate the effects of possible macroscopic coherence on 
radiative neutrino scattering on electrons. The corresponding cross 
sections with the net electron recoil momentum constrained by 
$|\vec{p}\,'|\le p_0$ with a small cutoff $p_0$ were found in 
Section~\ref{sec:neutrino}. 
Consider first radiative neutrino-electron scattering mediated by the usual 
NC and CC weak interactions. The constrained differential and integrated 
elementary cross sections are given in eqs.~(\ref{eq:WE4}) and 
(\ref{eq:WE5}), and are proportional to $p_0^4$. To find the cross section 
per one target electron one has to multipy these cross sections by the 
number of electrons in the coherent volume,
\be
N_{0e}\,\simeq\,n_e L_0^3 \,\simeq\, 
n_e\left(\frac{2\pi}{p_0}\right)^3\,,
\label{eq:N0e}
\ee  
where $n_e$ 
is the electron number density in the target. 
As a result, the cross sections per one target electron turn out to be 
proportional to $p_0$ and are maximized for maximal possible value of $p_0$, 
which corresponds to the absence of macrosopic coherence. What actually 
happens in this case is that macroscopic coherence can be achieved, but it 
requires such a stringent constraint on the value of $|\vec{p}\,'|$ (and so 
on the phase space available to the final-state particles) that the 
resulting cross sections are much smaller than those in the incoherent case. 
That is, macroscopic coherence is possible, but it leads to a reduction of 
the cross section rather than to its increase. 

The situation is different for neutrino magnetic (or electric) dipole moment 
mediated radiative neutrino scattering. As discussed in 
Section~\ref{sec:numagn}, for small $\vec{p}\,'$ the cross sections 
get an enhancement due to the propagator of the virtual photon being close to 
its mass-shell pole. The enhacement is, however, rather modest: the 
constrained elementary cross sections 
(\ref{eq:MM5})-(\ref{eq:MM7})
are proportional to $p_0^3$ rather than to $p_0^4$, as it was in the case 
of weak NC and CC mediated radiative process. As before, to obtain the 
cross sections per target electron we have to multiply the constrained 
elementary cross sections by $N_{0e}$ given by eq.~(\ref{eq:N0e}). 
The factor $p_0^3$ in the cross sections~(\ref{eq:MM5})-(\ref{eq:MM7}) then 
gets canceled by $1/p_0^3$ from eq.~(\ref{eq:N0e}), i.e.\ to leading order 
in the small $p_0$ the resulting cross sections per target electron are 
$p_0$-independent.%
\footnote{
As follows from the derivation of eqs.~(\ref{eq:MM5})-(\ref{eq:MM7}), 
this is correct only when $p_0$ satisfies 
$L^{-1}\lesssim p_0 \ll \omega,\, \omega_\gamma,\, \omega-\omega_\gamma$. 
}
{}From eqs.~(\ref{eq:MM5}) and (\ref{eq:MM7}) we then find 
\be
\frac{d\overline{\sigma}_m}{d\omega_\gamma}\simeq 
\frac{\mu_\nu^2 e^4}{6}\,
\frac{(\omega-\omega_\gamma)}{\omega\omega_\gamma^2}\,
\frac{n_e}{m_e^2}
\,,
\label{eq:MM5a}
\vspace*{1mm}
\ee
\be
\overline{\sigma}_m(\omega_\gamma>\omega_0)\simeq
\frac{1}{6}\,
\mu_\nu^2 e^4 
\frac{n_e}{m_e^2 \omega_0}=
\frac{8}{3}\pi^2\,
\frac{\mu_\nu^2 \alpha^2}{m_e^2 \omega_0}\,n_e
\,.
\vspace*{1mm}
\label{eq:MM7a}
\ee
Here the lines over $\sigma_m$ are to denote the cross sections per one target 
electron with coherence effects taken into account, and 
we have assumed $\omega_0\ll \omega$ in eq.~(\ref{eq:MM7a}).

The simplified approach we have adopted to evaluate the coherence effects, 
namely, to introduce the cutoff $p_0\sim 2\pi/L_0$ on the electron recoil 
momentum, replace the structure factors $F(\vec{k}-\vec{k}'-\vec{k}_\gamma)=
F(\vec{p}\,')$ within the coherence volume $L_0^3$ by unity and then 
multiply the obtained elementary cross sections by the number electrons in 
the coherent volume, actually proves to be rather accurate. 
As we shall see, it just slightly overestimates the numerical 
factors in the cross sections (\ref{eq:MM5a}) and (\ref{eq:MM7a}). 
A more accurate estimate is obtained if one notes that for a  macroscopically 
large number of electrons of the target contributing coherently to the cross 
section of the process, the summation in the expression for the 
structure factor in eq.~(\ref{eq:F4}) can be replaced by integration. 
This yields 
\be
F(\vec{k}-\vec{k}'-\vec{k}_\gamma)
\simeq\frac{N_e}{V} (2\pi)^3 \delta^3(\vec{k}-\vec{k}'-\vec{k}_\gamma)\,,\quad~
\label{eq:sumint1}
\ee
where $N_e=n_e V$ is the total electron number in the target and $V$ is the 
target's volume. Mutiplying the squared matrix element of the elementary 
process by 
\be
|F(\vec{k}-\vec{k}'-\vec{k}_\gamma)|^2 \simeq N_e n_e (2\pi)^3 
\delta^3(\vec{k}-\vec{k}'-\vec{k}_\gamma)\,,
\label{eq:sumint2}
\ee
performing the integration over the momenta of the scattered neutrino 
and the recoil electron as well as over the directions of the photon emission 
and dividing by $N_e$, for the differential cross section per one target 
electron we obtain 
\be
\frac{d\overline{\sigma}_m}{d\omega_\gamma}=
\frac{\mu_\nu^2 e^4}{4\pi}\,
\frac{(\omega-\omega_\gamma)}{\omega\omega_\gamma^2}\,
\frac{n_e}{m_e^2}
\,,
\label{eq:MM5b}
\vspace*{1mm}
\ee
which has the same structure as (\ref{eq:MM5a}), but is smaller by a factor 
of $3/2\pi\simeq 0.5$. The $\omega_\gamma$-integrated cross section in the 
limit $\omega_0\ll \omega$ will be smaller than the expression in 
eq.~(\ref{eq:MM7a}) by the same factor. Note that the same approach applied to 
the radiative neutrino-electron scattering mediated by the NC and CC weak 
interactions would yield vanishing cross sections of coherent scattering per 
target electron. This corresponds to 
the already discussed fact that in this case macroscopic 
coherence, though possible for sufficiently small electron recoils, would lead 
to vanishingly small cross sections. 
 
Can the coherent enhancement of the neutrino magnetic moment mediated radiative 
neutrino scattering help us to increase the experimental sensitivity to the 
neutrino magnetic dipole moments or even to detect them? 
The best laboratory limits on neutrino magnetic moments come from the 
experiments on elastic $\nu e$ scattering at reactors, 
where one looks for possible deviations of the measured differential cross 
section from the usual one mediated by the weak CC and NC processes. The cross 
section due to the $\mu_\nu$-induced elastic $\nu e$ scattering is 
\be
\frac{d\sigma_m^{el}}{dT}=
\frac{\mu_\nu^2 e^2}{4\pi}
\Big(\frac{1}{T}-\frac{1}{\omega}\Big)\simeq \frac{\mu_\nu^2\alpha}{T}\,,
\label{eq:elast}
\ee
where $T$ is the kinetic energy of the recoil electron and in the last 
(approximate) equality it is assumed that $T\ll \omega$.  
This can be compared with the differential cross section (\ref{eq:MM5b}) 
of the radiative neutrino-electron scattering, where a forward photon rather 
than the recoil electron is detected. For a numerical estimate, we set in 
eq.~(\ref{eq:MM5b}) 
\be
n_e=N_A \rho(\rm{g/cm^3})Y_e\;{\rm cm}^{-3}\simeq 
(1.33\;{\rm keV})^3 \,\rho(\rm{g/cm^3})\,,
\label{eq:ne}
\ee
where $N_A$ is the Avogadro constant, $\rho$ is the density of the target 
material, $Y_e$ is the number of electrons per nucleon in the target, 
and in the last (approximate) equality we have set $Y_e=1/2$. In the regime 
$\omega_\gamma\ll \omega$ this gives 
\be
\frac{d\overline{\sigma}_m}{d\omega_\gamma}\simeq 
4\pi\alpha^2 \mu_\nu^2\,
\frac{(1.33~{\rm keV})^3}{m_e^2\omega_\gamma^2}\,\rho({\rm g/cm^3})
\,.
\label{eq:MM5c}
\vspace*{1mm}
\ee
Taking for an estimate $\omega_\gamma\sim 10$ eV, which is about the 
smallest value that would allow to avoid the $\omega^4$ suppression of the 
radiative cross section, $\rho\sim 1$ g/cm$^3$, and comparing 
eqs.~(\ref{eq:elast}) and (\ref{eq:MM5c}), we find that even in the most 
optimistic case the cross section of coherently enhanced radiative 
neutrino-electron scattering exceeds that of the incoherent elastic 
scattering only for the electron recoil energies satisfying $T\gtrsim 100$ 
keV. At the same time, reactor experiments are currently probing $\nu e$ 
scattering in the sub-keV region of the recoil energies $T$, where the 
cross section of the incoherent elastic scattering dominates. 
Still, it should be noted that experimentally detecting $\sim 10$ -- 100 
eV photons may be easier than detecting electron recoil energies in the 
same range. 

A potentially important advantage of the coherent radiative 
$\mu_\nu$-mediated $\nu e$ scattering 
is that it could in principle allow detection of very low energy 
neutrinos. Consider, e.g., neutrinos of energy $\omega\sim 100$ eV. For 
the elastic $\nu e$ scattering the electron recoil energies would then be 
$T\le 2\omega^2/m_e\simeq 0.04$ eV, which is far too small to be measured 
in a foreseeable future. At the same time, detection of photons of energy 
$\sim 100$ eV which would be produced through coherent radiative $\nu e$ 
scattering does not pose any problem. However, the observability of the 
radiative process would 
depend crucially on the currently unknown values of the neutrino magnetic 
dipole moments (which, of course, applies to the elastic process as well). 
To give an idea of the magnitude of the expected cross section we rewrite 
below the expression for $d\overline{\sigma}_m/d\omega_\gamma$ 
given in eq.~(\ref{eq:MM5c}) in convenient units: 
\be
\frac{d\overline{\sigma}_m}{d\omega_\gamma}\simeq 
2.06\times 10^{-56}
\left(\frac{\mu_\nu}{10^{-12}\mu_B}\right)^2\,
\rho({\rm g/cm^3})\,\left(\frac{100~{\rm eV}}{\omega_\gamma}\right)^2\,
~{\rm cm^2/eV}
\,.
\label{eq:MM5d}
\vspace*{1mm}
\ee
Here $\mu_B=e/2m_e$ is the electron Bohr magneton. 

Turning now to radiative coherent axion-photon conversion on electrons
considered in Section~\ref{sec:axions}, we note that this process is similar
to the neutrino magnetic moment induced radiative $\nu e$ scattering in that
the interaction with electrons is mediated by the photon exchange and
the constrained elementary cross section of the process scales as $p_0^3$ at
small $p_0$. The assessment of the effects of possible macroscopic coherence
is therefore also similar to the one for $\mu_\nu$-mediated $\nu e$ scattering.
Following the procedure outlined at the beginning of this Section, for the 
cross section per target electron with coherence effects taken into account we 
find 
\be
\bar{\sigma}_a\simeq \frac{2}{3}\pi^2\,\frac{g_{a\gamma\gamma}^2\alpha^2}
{m_e^2\,\omega}n_e
\simeq \frac{2}{3}\, g_{a\gamma\gamma}^2\alpha^2 \pi^2 
\frac{(1.33\;{\rm keV})^3} {m_e^2\,\omega}\,\rho(\rm{g/cm^3})
\,.
\label{eq:ax4}
\ee
If instead one calculates the structure factor by replacing the summation 
by integration as in eq.~(\ref{eq:sumint1}), 
the result will differ from (\ref{eq:ax4}) by a factor of $3/2\pi$. 

It is instructive to compare this with other processes for axion
detection. The most relevant process for experimental searches for
relativistic axions is axion-photon conversion in an external magnetic
field which is used in searches for axions from the sun with
helioscopes. An axion traveling through a transverse magnetic field $B$
over a length $L$ is converted to photons with a probability $P$ given
by~\cite{VanBibber:1987rq}
\begin{align}
P= 2.4 \times 10^{-21} \left(g_{a\gamma \gamma}\times 10^{10} \, \mbox{GeV}
\right)^2 \left(\frac{B}{\mbox{T}} \right)^2 \left(\frac{L}{\mbox{m}} \right)^2
F\,,
\label{eq:magn1}
\end{align}
where the form factor  
\begin{align}
F=\left(\frac{2 \sin(\frac{q L}{2})}{q L}\right)^2
\label{eq:FF}
\end{align}
accounts for the loss of coherence as a function of the momentum
transfer $q$. For practical purposes $F\approx1$ is a good 
approximation for the energy transfers of interest here. The photon production 
rate is $\Gamma_a = j_a \cdot P \cdot A_{\rm eff}$, where $j_a$ is the
flux of axions from a given source and $A_{\rm eff}$ is the effective area
of the detector. Realistic values for recent axion helioscopes such as
CAST~\cite{Zioutas:2004hi} are $B\approx 10$ T, $L\approx 10$ m and
$A_{\rm eff} \approx 1\,\mbox{cm}^2$ and we expect 
\be
\Gamma_a \approx 2.4\times 10^{-17} \mbox{cm}^2 \,
(g_{a\gamma \gamma}\times 10^{10}\,\mbox{GeV})^2 j_a\,. 
\label{eq:magn2}
\ee
The photon production rate due to the coherent radiative axion-photon 
conversion mechanism considered in Section~\ref{sec:axions} is 
\be 
\Gamma_a = j_a \bar{\sigma}_a n_e V\simeq  \bar{\sigma}_a 
(1.33\;{\rm keV})^3 \rho(\mbox{g/cm}^3) V j_a\,,
\label{eq:evRate1}
\ee
where $V$ denotes the volume of the detector, and the cross section per target 
electron $\bar{\sigma}_a$ was given in eq.~(\ref{eq:ax4}).  
We can now compare the photon flux due to conversion in a magnetic field
with the flux from radiative scattering on electrons. 
Taking for an estimate $\omega \simeq 3$ keV, the characteristic energy of 
axions produced in the sun \cite{vanBibber:1988ge}, and 
$\rho\simeq 1\,\mbox{g/cm}^3$, we find that the rate in eq.~(\ref{eq:evRate1}) 
is by far small compared to the magnetic conversion rate~(\ref{eq:magn2}) 
for all reasonable detector volumes.  
Therefore, macroscopic coherence of the radiative axion conversion on 
electrons is not competitive with the coherent conversion in a magnetic field.

\section{\label{sec:disc} Summary and discussion}

We have considered the possibility of achieving macroscopic coherence 
in neutrino detection experiments. For the elastic neutrino 
scattering processes, coherence at macroscopic scales can only be 
attained at the expense of unmeasurably small recoil energies of the 
target particles, $E_{rec}\sim 10^{-43}$~eV, and so is of no use for 
neutrino detection. We therefore concentrated on radiative 
neutrino scattering on electrons $\nu e\to \nu e\gamma$, which has a 
number of attractive features:

\begin{itemize}

\item
Unlike the elastic neutrino-nucleus scattering, elastic and radiative $\nu e$ 
scattering processes are sensitive to neutrino flavour and so could serve for 
studying neutrino flavour oscillations.

\item 
In the case of radiative scattering the emitted photon rather than the recoil 
electron can be detected. As the photon energy $\omega_\gamma$ 
can be as large as the momentum $|\vec{q}|$ transferred to the electron from 
the neutrino, the process is sensitive to the neutrino momentum 
transfer rather than to the (very small) recoil energy of the target electron. 

\item
An important advantage of the radiative $\nu e$ scattering is that neither 
the momentum transfer $|\vec{q}|$ 
nor the photon energy $\omega_\gamma$ need to be small in order to ensure 
macroscopic coherence of the process. What actually has to be small 
is the net recoil momentum of the target electron, which is the 
difference between the momentum transfer from the neutrinos 
$\vec{q}=\vec{k}-\vec{k'}$ and the momentum $\vec{k}_\gamma$ carried away by 
the photon. This can happen even when both $|\vec{q}|$ and 
$|\vec{k}_\gamma|=\omega_\gamma$ are large compared to the inverse linear 
size of the target (or of a macroscopic volume within the target). 

\end{itemize} 

The drawback is that the requirement 
$\vec{k}-\vec{k'}\simeq \vec{k}_\gamma$ puts 
a stringent constraint on the kinematics of the process, reducing the  
phase space accessible to the final-state particles and in general also 
affecting  the dynamics of the process. 
This leads to a suppression of the cross section of the elementary process 
of neutrino radiative scattering on a single target electron. 
We have found that for the usual NC and CC induced $\nu e$ interactions, 
macroscopic enhancement of the number of electrons contributing coherently 
to the total cross section for small electron recoil momenta  
cannot overcome the suppression of the elementary cross section, and the net 
effect is a strong reduction of the total cross section compared to the 
incoherent case. 

The situation is different for the radiative $\nu e$ scattering induced by 
neutrino magnetic (or electric) dipole moments. In that case the amplitude 
of the process is dynamically enhanced for $\vec{k}-\vec{k'}\simeq 
\vec{k}_\gamma$ because of the propagator of the virtual photon being close 
to its mass-shell pole. The suppression of the elementary cross section due 
to the decrease of the phase space volume at small electron recoil momenta 
is then compensated by the macroscopically large number of electrons 
contributing coherently to the photon production rate. However, the cross 
section of the radiative process has some additional small factors (such as 
an extra power of $\alpha$) compared to the non-radiative one. At the same 
time, the usual increase of the radiative cross section at small photon 
energies is limited by the requirement that $\omega_\gamma$ exceed the 
characteristic atomic frequencies $\omega_{at}$ for neutrino scattering on 
atomic electrons or plasma frequency $\omega_p$ for scattering on free 
electrons in a conductor. As a consequence, even in the most optimistic case 
($\omega_\gamma$ close to its lower limit) for coherently enhanced radiative 
neutrino scattering the cross section per target electron exceeds the 
usual differential cross section of $\mu_\nu$-mediated $\nu e$ scattering 
only when in the latter case the kinetic energies of recoil electrons 
satisfies $T\gtrsim 100$ keV. In any case, no increase of the experimental 
detection rates by a huge factor, which could be expected for a 
macroscopically coherent process, takes place.

The $\mu_\nu$-mediated coherent radiative $\nu e$ scattering has another 
advantage, though: it allows in principle to detect neutrinos in the energy 
domain $\sim$100 eV -- a few keV, which is currently not accessible to the 
experiment. Possible sources of such neutrinos include nuclear reactors, 
the sun and relic supernovae, whose neutrino spectrum can be softened by 
large redshifts. At the moment, the corresponding expected fluxes are 
essentially unknown, except for solar neutrinos, for which only 
one calculation in the keV energy range exists \cite{vitagliano}.  
To detect keV-range neutrinos through the usual elastic 
$\mu_\nu$-mediated $\nu e$ scattering, one would need to measure the 
electron recoil energies on the order of $\sim 1$ eV, whereas the 
current sensitivity is at the level of $\sim 0.3$ keV. At the same 
time, detecting a 100 eV -- a few keV photon does not pose any 
experimental problem. Obviously, whether or not such a detection of 
very low energy neutrinos will ever become possible depends crucially 
on the (currently unknown) values of the neutrino magnetic or 
electric dipole moments. This applies, of course, to both the 
radiative scattering discussed here and the usual elastic $\nu e$ 
scattering.

To summarize very briefly our findings, the elastic and radiative neutrino 
scattering processes that we have considered do not allow strong 
increase of neutrino detection cross sections through macroscopic 
coherence. For elastic scattering, the cross section per target particle 
can be increased by a huge factor $\sim N^{1/3}$, where $N$ is the total 
number of scatterers in the target; however, in this case macroscopic 
coherence requires neutrino scattering in practically forward direction with 
essentially zero momentum transfer and so with no observable signatures. 

In the case of coherent radiative scattering, the emitted photons can 
in principle be easily detected, giving a clear experimental 
signature; however, the constraints on the kinematics of the process 
coming from the requirement of macroscopic coherence lead to very 
small cross sections per target particle. The only exception may be 
radiative scattering mediated by neutrino magnetic or electric dipole 
moments, but the experimental prospects for such processes are 
unclear because of the the unknown neutrino electromagnetic moments.

Does all this mean that macroscopically coherent detection of 
neutrinos is not possible in principle? We did not prove this as a 
theorem, but we believe that with our studies the observability of such 
a coherent enhancement becomes increasingly less probable.

We have also considered radiative axion photon conversion on
electrons. Similarly to the case of $\mu_\nu$-mediated coherent radiative 
$\nu e$ scattering, the amplitude receives a dynamical enhancement 
in the kinematic regime of interest to us since the photon propagator goes 
almost on shell. Therefore, the decrease of the phase space volume is again 
to some extent compensated by the contribution of a macroscopic number of 
electrons. However, in this case the radiative conversion process should be
compared with the most relevant detection process employed by ongoing
experiments, i.e. the axion-photon conversion in an external magnetic 
field (the inverse Primakoff effect). This process 
is already macroscopically coherent, and it turns out that, despite
the coherent enhancement, the radiative axion photon conversion studied in this 
paper is not competitive in practice. 

Can macroscopically coherent radiative scattering processes be employed for 
detecting dark matter particles? We have shown that for the conventionally 
discussed non-relativistic dark matter particle 
candidates this is not possible. The coherent enhancement mechanism 
studied here may, however, work for the detection of relativistic 
particles which usually exist in multi-component dark matter models, 
an example being the ``boosted'' dark matter \cite{boosted}. We 
believe that such a possibility deserves a dedicated study.

\vspace*{1.5mm}
{\em Acknowledgments.} The authors are grateful to Werner Maneschg and 
Alexei Smirnov for useful discussions. 

\appendix
\renewcommand{\theequation}{\thesection\arabic{equation}}
\appsection
\renewcommand{\thesection}{\Alph{section}}
\section*{Appendix \Alph{section}: Kinematics of the process}
Consider the kinematics of the process 
\be
X(k)+e(p)\to X(k')+e(p')+\gamma(k_\gamma)\,,
\label{eq:proc2} 
\ee
where $X$ is 
a projectile particle of mass $M$. 
In the rest frame of the initial-state electron the 4-momenta of the incident 
$X$-particle, initial-state electron, scattered 
$X$-particle, final-state electron and emitted photon are, 
respectively, 
\be
k=\big(\omega, \vec{k}\big)\,,\quad p=(m_e, \vec{0})\,,\quad
k'=\big(\omega', \vec{k}'\big)\,,\quad p'=\big(E_{p'}, \vec{p}\,'\big)\,,
\quad k_\gamma=\big(\omega_\gamma, \vec{k}_\gamma\big)\,,
\label{eq:4mom1}
\ee
where 
\be
\omega=\sqrt{\vec{k}\,^2+M^2},\quad 
\omega'=\sqrt{\vec{k}'^2+M^2}\,,\quad 
E_{p'}=\sqrt{\vec{p}\,'^2+m_e^2}\,,\quad 
\omega_\gamma=|\vec{k}_\gamma|\,.
\label{eq:energies}
\ee
The energy and momentum conservation laws yield 
\be
\omega=\omega'+(E_{p'}-m_e)+\omega_\gamma\,, 
\label{eq:encons1}
\ee
\be
\vec{k}=\vec{k}'+\vec{p}\,'+\vec{k}_\gamma\,.\qquad\quad~~
\label{eq:momcons1}
\vspace*{1.0mm}
\ee
In what follows we will be assuming the recoil electron to be non-relativistic 
and will neglect its kinetic energy. The energy conservation condition then 
simplifies to 
\be
\omega=\omega'+\omega_\gamma\,. 
\label{eq:encons1a}
\ee
Expressing $\vec{k}'$ from eq.~(\ref{eq:momcons1}) and substituting it into 
(\ref{eq:encons1a}), we get 
\be
(\vec{k}-\vec{k}_\gamma-\vec{p}\,')^2=\vec{k}^2-2\omega_\gamma
\sqrt{\vec{k}\,^2+M^2}+\omega_\gamma^2\,,
\label{eq:relat1}
\ee
or  
\be
\vec{p}\,'^2-2|\vec{p}\,'|R\cos\theta_{\vec{p}'(\vec{k}-\vec{k}')}
-2|\vec{k}|\omega_\gamma\cos\theta_\gamma=-2\omega_\gamma\sqrt{\vec{k}^2+M^2}
\,. 
\label{eq:relat2}
\vspace*{1mm}
\ee
Here $\theta_\gamma$ is the angle between $\vec{k}_\gamma$ and $\vec{k}$, 
\,$\theta_{\vec{p}'(\vec{k}-\vec{k}')}$ is the angle between 
$\vec{p}\,'$ and $\vec{k}-\vec{k}_\gamma$, and 
\be
R\equiv |\vec{k}-\vec{k}_\gamma|=\sqrt{\vec{k}^2+\omega_\gamma^2
-2\omega_\gamma|\vec{k}|\cos\theta_\gamma}\,.
\label{eq:relat3}
\ee
We will be interested in the regime of very small $|\vec{p}\,'|$. Let us first 
demonstrate that for $M\neq 0$ the quantity $|\vec{p}\,'|$ cannot be 
arbitrarily small. Indeed, in the limit $\vec{p}\,'\to 0$ eq.~(\ref{eq:relat2})
leads to unphysical 
$\cos\theta_\gamma=\sqrt{\vec{k}\,^2+M^2}/|\vec{k}|>1$. 
Next, we assume $|\vec{p}\,'|$ to be non-zero but small, such that the 
$\vec{p}\,'^2$ term in (\ref{eq:relat2}) can be neglected. We should then 
also replace $\cos\theta_\gamma$ in the factor $R$ by unity, i.e.\ 
set $R=||\vec{k}|-\omega_\gamma|$. This is because $R$ enters 
in eq.~(\ref{eq:relat2}) multiplied by $|\vec{p}\,'|$, and for small 
$|\vec{p}\,'|$ the difference $1-\cos\theta_\gamma={\cal O}(|\vec{p}\,'|)$; 
thus, keeping $1-\cos\theta_\gamma$ in $R$ would lead to terms of higher order 
of smallness in $|\vec{p}\,'|$. 
Requiring $\cos\theta_\gamma\le 1$, we then find from eq.~(\ref{eq:relat2}) 
\be
\sqrt{\vec{k}^2+M^2}-|\vec{k}|
\;\le\; 
\frac{||\vec{k}|-\omega_\gamma|}{\omega_\gamma}\cdot
|\vec{p}\,'|\cos\theta_{\vec{p}'(\vec{k}-\vec{k}')}\,.
\label{eq:relat4}
\ee
For non-relativistic 
projectile particles ($|\vec{k}|\ll M)$, eq.~(\ref{eq:relat4}) yields 
\be
M-\frac{|\vec{k}|}{\omega_\gamma}\cdot
|\vec{p}\,'|\cos\theta_{\vec{p}'(\vec{k}-\vec{k}')}\,\le\, |\vec{k}|\,,
\label{eq:Mbound1}
\ee
which means that the two terms on the left hand side should nearly cancel each 
other. This requires 
\be
\omega_\gamma\simeq\frac{|\vec{k}|}{M}
|\vec{p}\,'|\cos\theta_{\vec{p}'(\vec{k}-\vec{k}')}\,\ll\,|\vec{p}\,'|\,.
\label{eq:Mbound2}
\ee
As discussed in Section~\ref{sec:macrcoh}, to achieve macroscopic coherence 
one needs $|\vec{p}\,'|\lesssim 10^{-5}$ eV; condition (\ref{eq:Mbound2}) then 
implies that the requirement $\omega_\gamma \gtrsim \omega_{at}$ 
(or $\omega_\gamma \gtrsim \omega_{p}$ for scattering on free electrons in 
a conductor) is badly violated 
for non-relativistic projectiles, leading to a strong suppression of the 
cross section of process (\ref{eq:proc2}). Similar estimates apply and the 
same conclusion holds for the case of moderately relativistic 
projectiles.%
\footnote{It is convenient to write in this case $\sqrt{\vec{k}^2+M^2}=
|\vec{k}|(1+a)$, where $a={\cal O}(1)$. Noting that the factor 
$[|\vec{k}|-\omega_\gamma]/\omega_\gamma$ is maximised for 
$\omega_\gamma\ll |\vec{k}|$, we find from (\ref{eq:relat4}) $a\omega_\gamma 
\le|\vec{p}\,'|\cos\theta_{\vec{p}'(\vec{k}-\vec{k}')}\ll \omega_{at}$.  
}

In the ultra-relativistic regime $|\vec{k}|\gg M$ eq.~(\ref{eq:relat4}) 
yields an upper bound on $M$:
\be
M^2\,\le\, 2\omega
\frac{\omega-\omega_\gamma}{\omega_\gamma}
|\vec{p}\,'|\cos\theta_{\vec{p}'(\vec{k}-\vec{k}')} \,.
\label{eq:Mbound3}
\ee
For $\omega_\gamma$ not too close to its upper limit $\omega$ and 
$\cos\theta_{\vec{p}\,'(\vec{k}-\vec{k}')}\sim 1$ eq.~(\ref{eq:Mbound3}) 
yields $M^2\,\lesssim 2\omega|\vec{p}\,'|$.  
For $|\vec{p}\,'|\sim 10^{-5}$ eV and $\omega\sim 1$ keV this gives 
$M\lesssim 0.14$ eV, which can be readily satisfied when the projectiles are 
neutrinos or axions.  

\appsection
\renewcommand{\thesection}{\Alph{section}}
\section*{Appendix \Alph{section}: 3-body phase space volume}

We shall now consider the regime of relativistic projectiles assuming that 
condition (\ref{eq:Mbound3}) is satisfied with a large margin. This will 
allow us to treat the projectile as essentially massless. 

Consider the 3-body phase space volume integral
\be
R_3\,\equiv 
\int
\frac{d^3p'}{2E_{p'}}\frac{d^3k_\gamma}{2\omega_\gamma}
\frac{d^3k'}{2\omega'}\delta^3(\vec{k}-
\vec{k}'-\vec{p}\,'-\vec{k}_\gamma)\delta(\omega-\omega'-\omega_\gamma)\,. 
\label{eq:phspace1}
\ee 
in the limit of non-relativistic energies of the recoil electrons. 
We will calculate it in two cases: (i)~without additionally constraining 
$|\vec{p}\,'|$ and (ii) assuming that $|\vec{p}\,'|$ is limited from above 
by a small value $p_0$. A similar approach is used in the computations of 
the cross sections given in Section~\ref{sec:neutrino}. 

We start by finding $R_3$ without additionally constraining the 
electron recoil momentum $\vec{p}\,'$. It is convenient to first integrate 
over $\vec{p}\,'$ by making use of the $\delta^3$-function enforcing 
3-momentum conservation. 
A straightforward calculation then yields 
\be 
R_3=\frac{\pi^2\omega^3}{3m_e}. 
\label{eq:phspace2} 
\ee 

Next, we will calculate $R_3$ using a different integration order, 
which will be more convenient for studying the case of constrained 
$|\vec{p}\,'|$. To this end, we use the $\delta^3$-function to 
integrate over the momentum $\vec{k}'$ of the scattered projectile. 
One then has to substitute
 \be 
\vec{k}'=\vec{k}-\vec{p}\,'-\vec{k}_\gamma\, 
\label{eq:vec1} 
\ee 
in the integrand of the remaining integral. Using 
$\delta(\omega-\omega'-\omega_\gamma)= 
2\omega'\delta[(\omega-\omega_\gamma)^2-{\omega'}^2]$ and 
${\omega'}^{2}=\vec{k}'^{2}=(\vec{k}-\vec{p}\,'-\vec{k}_\gamma)^2$, 
we find 
\be 
\delta[(\omega-\omega_\gamma)^2-{\omega'}^2]\,=\,\delta\big( 
2|\vec{p}\,'|R\cos\theta_{\vec{p}\,'(\vec{k}-\vec{k}_\gamma)}-\vec{p}\,'^2 
-2\omega\omega_\gamma(1-\cos\theta_\gamma)\big)\,, 
\label{eq:deltaf1} 
\ee 
where $\theta_\gamma$ is the angle between $\vec{k}_\gamma$ and $\vec{k}$ and 
\be 
R\equiv|\vec{k}-\vec{k}_\gamma|=\sqrt{(\omega-\omega_\gamma)^2+2\omega 
\omega_\gamma x_\gamma}\,,\qquad x_\gamma\equiv 1-\cos\theta_\gamma\,. 
\label{eq:R} 
\ee 
Requiring $\cos\theta_{\vec{p}\,'(\vec{k}-\vec{k}_\gamma)}\le1$, we find that 
for fixed $\omega$ and $x_\gamma$ the quantity $|\vec{p}\,'|$ must lie in 
the interval $[p_{min}'\,,\,p_{max}']$, \vspace*{-1mm} where 
\be 
p_{min}'\,=\,R-(\omega-\omega_\gamma)\,,\qquad 
p_{max}'\,=\,R+(\omega-\omega_\gamma)\,. 
\vspace*{1mm}
\label{eq:allowed2} 
\ee 

We shall now consider the case when $|\vec{p}\,'|$ is limited from above by 
a value $p_0< p_{max}'$. The integration over $|\vec{p}\,'|$ is then done 
in the interval 
$|\vec{p}\,'|\in [p_{min}',\, p_0]$. From the condition $p_0>p_{min'}$ we  
find that $\cos\theta_\gamma$ must 
be in the interval given in eq.~(\ref{eq:allowed1}). 
Small $p_0$ therefore means that the photon is emitted in a nearly forward 
direction with respect to the incident projectile. From the energy-momentum 
conservation relations (\ref{eq:momcons1}) and (\ref{eq:encons1}) and the fact 
that we consider ultra-relativistic projectiles it follows that the same 
is true for the scattered projectile particle, i.e.\ for small $p_0$ its 
momentum $\vec{k}'$ is also nearly parallel to $\vec{k}$.

Note that in general $0\le 1-\cos\theta_\gamma\le 2$. Therefore, the condition 
$|\vec{p}\,'|\le p_0$ puts a non-trivial constraint on $x_\gamma=1-
\cos\theta_\gamma$ only when the expression on the right hand side of 
eq.~(\ref{eq:allowed1}) is smaller than 2. This yields  $p_0/2<\omega_\gamma$. 
On the other hand, the constraint $|\vec{p}\,'|\le p_0$ is only non-trivial 
when $p_0<p_{max}'$ for all $x_\gamma$. Setting $x_\gamma=0$ (which minimizes 
$p_{\max}'$ for a given $\omega_\gamma$) then yields $\omega_\gamma \le 
\omega-p_0/2$. Thus, in the constrained case under consideration 
the allowed range for the photon energy is 
\be
p_0/2\, \le\, \omega_\gamma \le \omega-p_0/2\,.
\label{eq:allowed3}
\ee
Since we are interested in tiny values of $p_0$, for all practical purposes 
the interval (\ref{eq:allowed3}) can be replaced by the usual allowed range 
for the photon energy, $0\le\, \omega_\gamma \le \omega$. 

Performing in (\ref{eq:phspace1}) the integration over $\vec{k}'$ by making 
use of the $\delta^3$ function, then integrating over the directions of the 
vector $\vec{p}\,'$ with the help of (\ref{eq:deltaf1}), over the modulus 
of this vector in the interval $[p_{min},\,p_0]$ and finally over 
$\vec{k}_\gamma$ (taking eq.~(\ref{eq:allowed2} into account), we find
\be 
R_3=\frac{\pi^2 p_0^3}{6m_e}\,. 
\label{eq:phspace3} 
\ee
This has to be compared with the unconstrained result (\ref{eq:phspace2}). 

\appsection
\renewcommand{\thesection}{\Alph{section}}
\section*{Appendix \Alph{section}: Squared matrix elements}

In this Appendix we collect the expressions for the squared moduli of the 
transition matrix elements $\overline{|{\cal M}|^2}$ for the processes 
considered in Sections \ref{sec:neutrino} and \ref{sec:axions}. 
Here, as usual, the line over $|{\cal M}|^2$ denotes the summation over the 
polarisations of the final particles and
averaging over the polarisations of the initial-state ones. 

For weak NC and CC induced radiative neutrino scattering, the calculations 
are most easily done in the Coulomb gauge. From eq.~(\ref{eq:M1}) one finds  
\[
\overline{|{\cal M}_w|^2}=
\frac{G_F^2 g_V^2 e^2}{2}\,
32\left\{\frac{1}{\omega_\gamma^2}
\bigg[{\vec{p}\,'}^2-\frac{(\vec{p}\,'\vec{k}_\gamma)^2}{\omega_\gamma^2}\bigg]
(\omega\omega'+\vec{k}\vec{k}\,') +2\bigg(\omega\omega'-\frac{
(\vec{k}\vec{k}_\gamma)(\vec{k}\,'\vec{k}_\gamma)}{\omega_\gamma^2}\bigg)
\right.
\hspace*{0.8cm}~~~~~~~~
\]
\be
\hspace*{1.0cm}~~~~~~~~
\left.-\frac{2}{\omega_\gamma}\bigg[
\omega\bigg(\vec{p}\,'\vec{k}\,'-\frac{
(\vec{p}\,'\vec{k}_\gamma) (\vec{k}\,'\vec{k}_\gamma)}{\omega_\gamma^2}\bigg)   
+\omega'\bigg(\vec{p}\,'\vec{k}-\frac{
(\vec{p}\,'\vec{k}_\gamma) (\vec{k}\vec{k}_\gamma)}{\omega_\gamma^2}\bigg)   
\bigg]
\right\}.
\label{eq:Msq1}
\ee

For neutrino magnetic (or electric) dipole moment induced radiative scattering, 
Lorentz gauge proves to be more convenient because it allows one to more 
easily get rid of angle-dependent denominators in most terms and thus to 
simplify the subsequent angular integrations. From eq.~(\ref{eq:M2}) we obtain 
\be 
\overline{|{\cal M}_m|^2}=\frac{\mu_\nu^2 e^4}{(-2k k')^2}\,8(k k')
\left\{2\bigg[
\frac{
[p(k+k\,')][p'(k+k')]}{
(p k_\gamma) 
(p'k_\gamma)} - 1\bigg](k k')-m_e^2\bigg[
\frac{p(k+k')}{p' k_\gamma}
-\frac{p'(k+k')}{p k_\gamma}
\bigg]^2
\right\}.
\label{eq:Msq2}
\ee

Calcualations for radiative axion-photon conversion are also more easily done 
in the Lorentz gauge. Since the complete expression for the 
squared matrix element is quite lengthy in that case (mostly due to the 
interference of the two parts of the amplitudes corresponding to the 
interchange of the 4-momenta of the two photons in the final state), we give 
here only the expression in the limit of small $|\vec{p}\,'|$ that is of main 
interest to us:  
\be
\overline{|{\cal M}_a|^2}\simeq 4g_{a\gamma\gamma}^2 e^4 \left(
\frac{\omega_1}{\omega_2}+\frac{\omega_2}{\omega_1}\right).
\label{eq:Msq3}
\ee

\end{document}